\documentclass[preprint,12pt]{elsarticle}

\usepackage{graphicx}
\usepackage{cases}
\usepackage{amssymb}
\usepackage{amsthm}

\newtheorem{resultat}{Result}
\newcommand{\ds}{\displaystyle}
\newcommand{\ts}{\textstyle}

\usepackage{color}
\definecolor{rouge}{rgb}{1,0,0}
\definecolor{bleu}{rgb}{0,0,1}
\definecolor{vert}{rgb}{0,0.5,0}
        
\definecolor{mygreen}{rgb}{0.15,0.7,0.15}
\definecolor{myred}{rgb}{0.9,0.05,0.05}
\definecolor{myblue}{rgb}{0.0352,0.4981,0.6509}

\usepackage[normalem]{ulem}
\usepackage{stmaryrd} 

\graphicspath{{Figures/}}

\journal{Wave Motion}

\begin{document}

\begin{frontmatter}

\title{A two-way model for nonlinear acoustic waves in a non-uniform lattice of Helmholtz resonators}

\author[ENSTA]{Jean-Fran\c{c}ois Mercier}
\ead{jean-francois.mercier@ensta.fr}
\author[LMA]{Bruno Lombard}\corref{cor1}
\ead{lombard@lma.cnrs-mrs.fr}
\cortext[cor1]{Corresponding author. Tel.: +33 491 16 44 13.}
\address[ENSTA]{POEMS, CNRS UMR 7231 CNRS-INRIA-ENSTA, 91762 Palaiseau, France}
\address[LMA]{Aix Marseille Univ, CNRS, Centrale Marseille, LMA, Marseille, France}

\begin{abstract}
Propagation of high amplitude acoustic pulses is studied in a 1D waveguide connected to a lattice of Helmholtz resonators. An homogenized model has been proposed by Sugimoto (J. Fluid. Mech., \textbf{244} (1992)), taking into account both the nonlinear wave propagation and various mechanisms of dissipation. This model is extended here to take into account two important features: resonators of different strengths and back-scattering effects. An energy balance is obtained, and a numerical method is developed. A closer agreement is reached between numerical and experimental results. Numerical experiments are also proposed to highlight the effect of defects and of disorder.
\end{abstract}


\end{frontmatter}



\section{Introduction}

We are concerned here with the dynamics of nonlinear waves in lattices. This subject has stimulated researches in a wide range of areas, including the theory of solitons and the dynamics of discrete networks. Numerous studies have been led in electromagnetism and optics \cite{Kevrekidis11}, and many physical phenomena have been highlighted, such as discrete breathers \cite{Lazarides06,Boechler10,Feng07}, chaotic phenomena \cite{Grabowski90,Wan90}, dynamical multistability \cite{Wan89,Li96} and solitons or solitary waves \cite{Li88a,Kartashov11}. We focus more specifically on solitary waves, which have been first observed and studied for surface waves in shallow water \cite{Russell44}. The main feature of these waves is that they can propagate without change of shape and with a velocity depending of their amplitude \cite{Remoissenet99}. They have been studied in many physical systems, as in fluid dynamics, optics or plasma physics.
 
For elastic waves, numerous works have highlighted the existence of solitary waves in microstructured solids \cite{Engelbrecht07}, periodic structures such as lattices or crystals \cite{Chetverikov11,Hao01,Hess10}, elastic layers \cite{Kuznetsov09,Lomonosov08,Mayer08}, layered structures coated by film of soft material \cite{Kovalev02}, periodic chains of elastics beads \cite{Billy12,Daraio06,Moleron14} and uniform or inhomogeneous rods \cite{Dai02,Khusnutdinova08,Billy14}.

On the contrary, only a few works have dealt with acoustic solitary waves. Such a lack is mainly explained by the fact that the intrinsic dispersion of acoustic equations is too low to compete with the nonlinear effects, preventing from the occurrence of solitons. To observe these waves, geometrical dispersion must be introduced. It has been the object of the works of Sugimoto and his co-authors \cite{Sugimoto99,Sugimoto92,Sugimoto96,Sugimoto04}, where the propagation of nonlinear waves was considered in a tube connected to an array of Helmholtz resonators. A model incorporating both the nonlinear wave propagation in the tube and the nonlinear oscillations in the resonators has been proposed. Theoretical and experimental investigations have shown the existence of acoustic solitary waves \cite{Sugimoto99}. We have developed a numerical modeling, and we have successfully compared simulations with experimental data \cite{Richoux15,Lombard14}.

One fundamental assumption underlying Sugimoto's model is that all the resonators are the same, which allows to use a homogenization process. The drawback of this approach is that the reflection of an incident wave by a defect (for instance, a resonator different from the others) cannot be considered. Similarly, wave propagation across resonators with variable features cannot be investigated. However, the case of variable resonators is important when studying the influence of manufacturing defects or the influence of aging of the guiding device on the wave propagation. 

The aim of this paper is to remedy these limitations, by building a model predicting two-way propagation across variable resonators. For this purpose, we introduce a discrete description of the resonators. A consequence is that the requirement of a long wavelength is no longer necessary. Second, we allow the reflection of waves. Finally, we reformulate the viscothermal losses, which leads to a formulation suitable for an energy balance. The new model writes as a single system coupling three unknowns, the velocity of the right-going wave $u^+$, the velocity of the left-going wave $u^-$ and the total excess pressure in the resonators $p$, induced by both waves:
\begin{equation}
\left\{
\begin{array}{l}
\ds
\frac{\partial u}{\partial t}^+ + \frac{\partial}{\partial x} \left(a_0 u^+ + b \ds \frac{(u^+)^2}{2}\right) + \frac{c}{a_0} \frac{\partial^{1/2}}{\partial t^{1/2}}u^+ - d \frac{\partial^2 u}{\partial x^2}^+ =-e(1-2 m p )\ds \frac{\partial p}{\textstyle \partial t}, \\
[12pt]
\ds
\frac{\partial u}{\partial t}^- + \frac{\partial}{\partial x} \left(- a_0 u^- + b \ds \frac{(u^-)^2}{2} \right) + \frac{c}{a_0} \frac{\partial^{1/2}}{\partial t^{1/2}}u^- - d \frac{\partial^2 u}{\partial x^2}^- = +e(1-2 m p )\ds \frac{\partial p}{\partial t}, \\
[12pt]
\ds
\frac{\partial^2 p}{\partial t^2} + f \frac{\partial^{3/2}}{\partial t^{3/2}}p + g p - m\frac{\partial^2 (p^2)}{ \partial t^2} + n \left|\frac{\partial p}{\partial t}\right|\,\frac{\partial p}{\partial t} = h (u^+ - u^-).
\end{array}
\right.
\label{NewModel}
\end{equation}
The precise meaning of $u^\pm$, $p$ and the coefficients in (\ref{NewModel}) will be detailed along section \ref{SecModel}. Some of the coefficients incorporate the individual features of the resonators and vary with $x$. 

The paper is organized as follows. In section \ref{SecModel}, the general equations in the tube and in the resonators are given, and the new model (\ref{NewModel}) is derived. In section \ref{SecEDP}, a first-order formulation is followed: it allows to determine an energy balance and also to build a numerical scheme. In section \ref{SecResults}, comparisons with experimental data show that a closer agreement is obtained than with the original Sugimoto's model. Numerical simulations are performed to investigate the properties of the acoustic solitary waves, in particular the robustness to defects and to random disorder. Conclusions and prospects are drawn in section \ref{SecConclu}. Technical results are given in the appendices.


\section{The new model}\label{SecModel}

\begin{figure}[htbp]
\begin{center}
\includegraphics[scale=0.70]{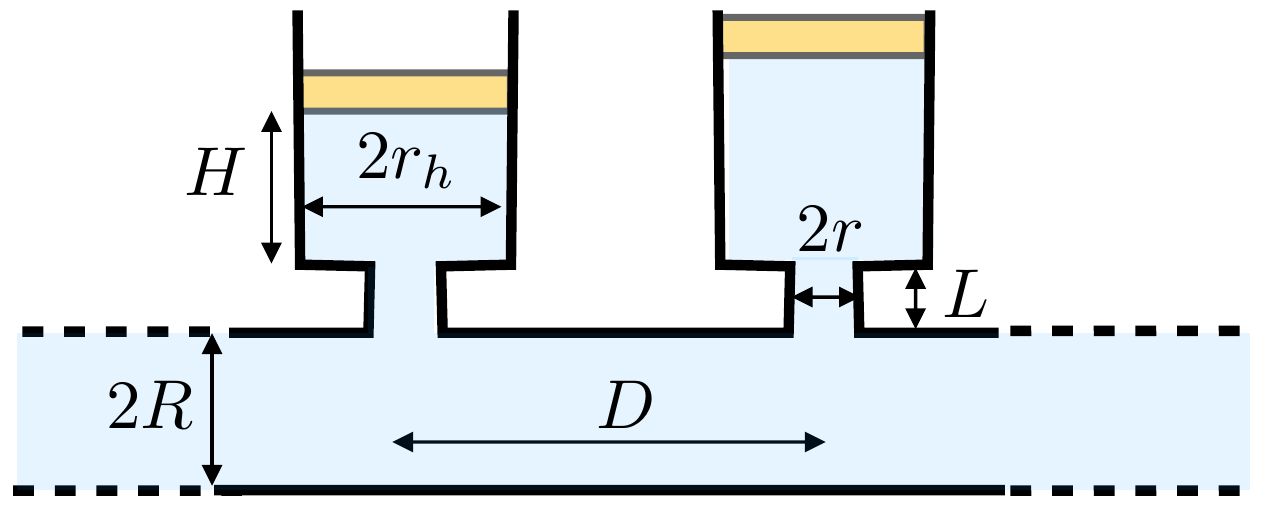}
\caption{\label{FigCrobar} sketch of the guide connected with Helmholtz resonators.}
\end{center}
\end{figure}

In this part, we derive Eq. (\ref{NewModel}).
The configuration under study is made up of an air-filled tube connected to uniformly distributed cylindrical Helmholtz resonators (see Fig. \ref{FigCrobar}). The geometrical parameters are the radius of the guide $R$, the axial spacing between successive resonators $D$, the radius of the neck $r$, the length of the neck $L$, the radius of the cavity $r_h$ and the height of the cavity $H$, which can vary from one resonator to the other. The cross-sectional area of the guide is $A=\pi\,R^2$ and that of the neck is $B=\pi\,r^2$, the volume of each resonator is $V=\pi\,r_h^2\,H$. 

\subsection{Equations in the tube}

\subsubsection{General equations}

Here we recall briefly the general equations governing the nonlinear acoustic waves in a tube \cite{Sugimoto92}. The physical parameters are the ratio of specific heats at constant pressure and volume $\gamma$; the pressure at equilibrium $p_0$; the density at equilibrium $\rho_0$; the Prandtl number Pr; the kinematic viscosity $\nu$; and the ratio of shear and bulk viscosities $\mu_v/\mu$. One defines the linear sound speed $a_0$, the coefficient of nonlinearity $b$, the sound diffusivity $\nu_d$ and the coefficient of attenuation $d$ by
\begin{equation}
a_0=\sqrt{\frac{\textstyle \gamma\,p_0}{\textstyle \rho_0}}, \qquad b=\frac{\ts \gamma+1}{\ts 2},\qquad
\nu_d=\nu\left(\frac{\textstyle 4}{\textstyle 3}+\frac{\textstyle \mu_v}{\textstyle \mu}+\frac{\textstyle \gamma-1}{\textstyle \mbox{Pr}}\right),\qquad  d=\frac{\ts \nu_d}{\ts 2}.
\label{}
\end{equation}
The characteristic angular frequency $\omega$ and the parameter of nonlinearity $\varepsilon$ are
\begin{equation}
\omega=\frac{2\,\pi\,a_0}{\lambda},\qquad \varepsilon=\frac{\gamma+1}{2}\frac{u}{a_0},
\end{equation}
where $\lambda$ is the typical wavelength imposed by the source, and $u$ is the amplitude of the gas velocity. The following assumptions are made:
\begin{itemize}
\item low-frequency regime ($\omega<\omega^*=\frac{1.84\,a_0}{R}$), so that only the plane mode propagates  \cite{Chaigne16};
\item weak acoustic nonlinearity: $\varepsilon \ll 1$ \cite{Hamilton98}.
\end{itemize}
Thanks to the first assumption, we can restrict to unidirectional propagation. Under the second assumption of weak nonlinearities, we are allowed to use Riemann invariants along the characteristics, noted with the $\pm$ subscript according to the way of propagation. Then, starting from the compressible Navier-Stokes equations, the flow in the tube can be described by the following nonlinear equation for the horizontal velocities $u^\pm$ \cite{Sugimoto92}:
\begin{equation} \label{invariant} \frac{\partial u}{\partial t}^\pm + \left( \pm a_0 + b u^\pm \right) \frac{\partial u}{\partial x}^\pm = \pm \frac{a_0}{2 \rho_0} F^\pm + d \frac{\partial^2 u}{\partial x^2}^\pm, 
\end{equation}
with
\begin{equation} 
F^\pm(x)=\frac{1}{A} \int_{\sigma(x)} (\rho v_n^\pm)(x,y(s),z(s)) ds.
\label{F} 
\end{equation}
The diffusivity of sound in the tube is introduced by the coefficient $d$. 

\begin{figure}[htbp]
\begin{center}
\includegraphics[width=8cm,clip]{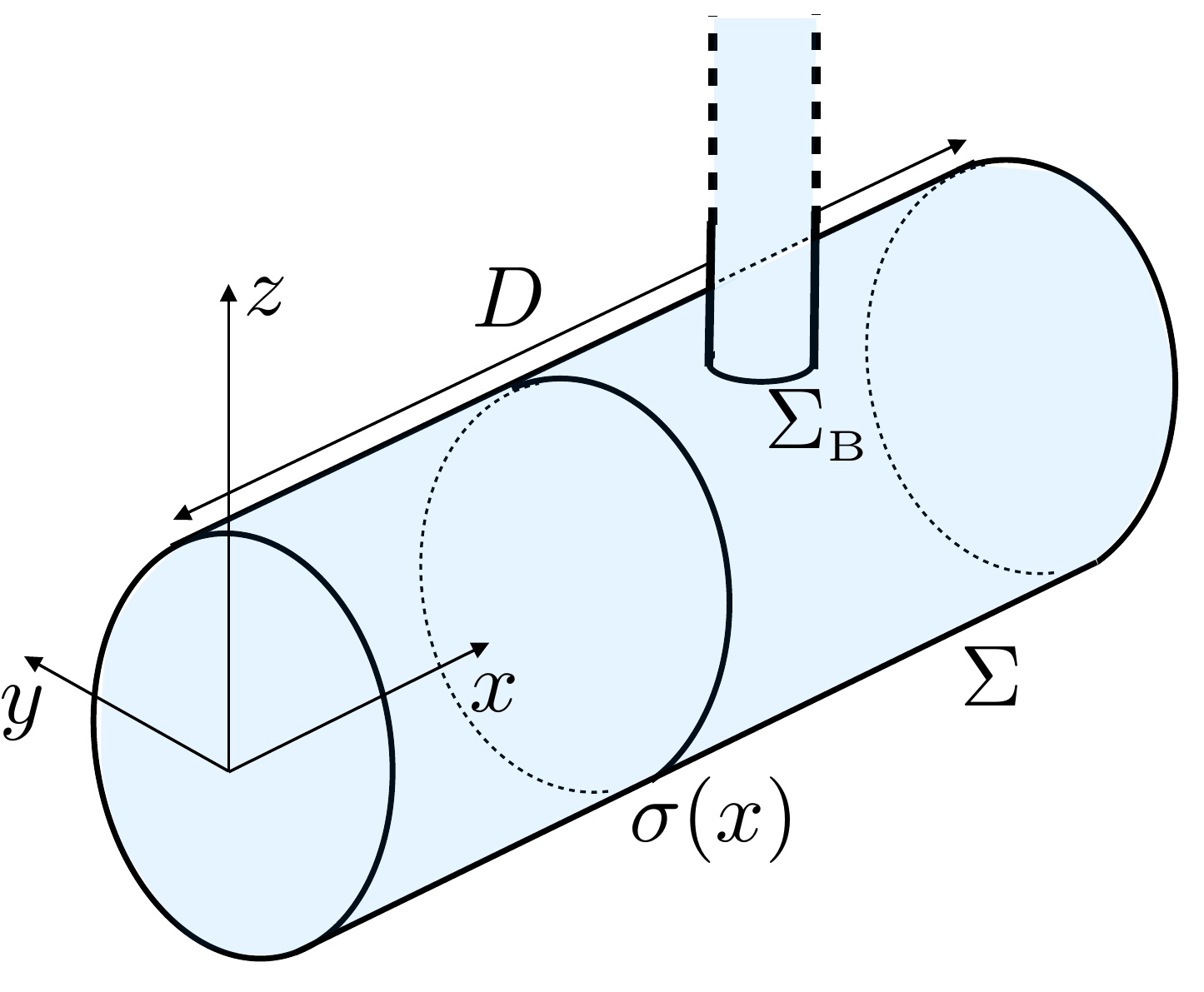}
\caption{\label{perspective} perspective view of the tube. The unit portion $\Sigma$ of the tube is connected to one resonator through the surface $\Sigma_B$. $\sigma$ is a path along the boundary of a cross section of the tube.}
\end{center}
\end{figure}

\begin{figure}[htbp]
\begin{center}
\includegraphics[scale=0.55]{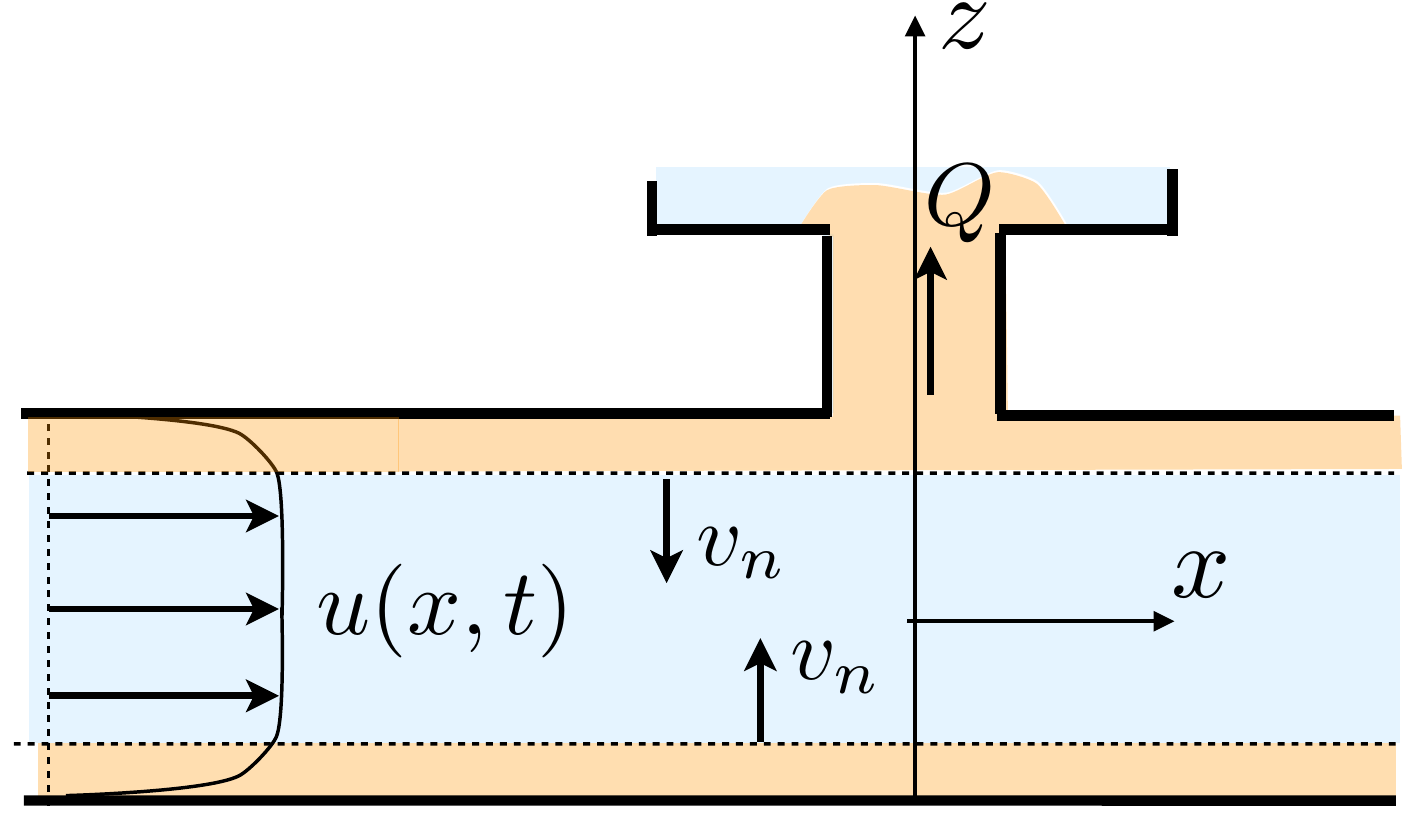}
\caption{\label{Figflux} definition of $v_n^\pm$. In beige is represented the viscous boundary layer and the inside of the resonators. $v_n^\pm$ is defined on the boundary of this beige area.}
\end{center}
\end{figure}

$F^\pm$ is the mean mass flux at position $x$, with $v_n^\pm$ the small deviation of the component of the velocity normal inward to the inner surface of the tube ($v_n^\pm$ lies in the $y-z$ plane). The integral is taken along the periphery $\sigma(x)$ of the cross-section at $x$ (see Fig. \ref{perspective}) with $s$ the curvilinear abscissa. $F^\pm$ is due to the coupling between the main flow $u^\pm$ and two areas, represented in beige on Fig. \ref{Figflux}. It takes into account the connections of the main tube with the resonators, and the wall friction due to the presence of a viscous boundary layer on the boundary of the tube.

We note $\Sigma$ a part of length $D$ (thus connected to only one resonator) of the tube, represented in Fig. \ref{perspective}. We note $\Sigma_B$ the part of $\Sigma$ connected to the cylindrical resonator, of area $B$. For simplicity, the junction between a resonator and the main tube is approximated by as a disc.
The real shape is more complicated: it is the intersection of two perpendicular tubes of different diameters. It follows \cite{Sugimoto92}:
\begin{equation} \label{A}
\begin{array}{rcl}
\rho v_n^\pm &=& - Q \quad \mbox{on} \quad \Sigma_B,\\[8pt]
 &=& \rho_0 v_b^\pm \quad \mbox{on} \quad \Sigma \backslash \Sigma_B,
\end{array}
\end{equation}
with
\begin{equation} \label{B}
\begin{array}{rcl}
Q &=& \displaystyle \frac{\textstyle V}{a_0^2 \,B} \left( 1 - 2 m p \right) \frac{\textstyle \partial p}{\textstyle \partial t},\\
v_b^\pm &=& C \displaystyle \sqrt{\nu} \frac{\textstyle \partial^{-1/2}}{\textstyle \partial t^{-1/2}}\frac{\textstyle \partial u}{\textstyle \partial x} ^\pm,
\end{array}
\end{equation}
where 
\begin{equation} \label{m}
m = \frac{\textstyle \gamma-1}{\textstyle 2\,\gamma\,p_0}.
\end{equation}
$Q$ is the mass flux density over $B$ into the resonator cavity (see Fig. \ref{perspective}). The reader is referred to the sections 2-1 and 2-2 of \cite{Sugimoto92} for a detailed derivation of (\ref{B}). The expression of $Q$ follows from the conservation laws in the throat and from the adiabatic relation in the cavity, up to the second order.  

In (\ref{B}), $v_b^\pm$ is the velocity at the edge of the viscothermal boundary layer. It involves a fractional operator of order -1/2, thus proportional to $1/(i\,\omega)^{1/2}$ in the frequency domain \cite{Chester64}. $C$ is the dissipation in the boundary layer:
\begin{equation}
C=1+\frac{\textstyle \gamma-1}{\textstyle \sqrt{\mbox{Pr}}}.
\end{equation}
These expressions are obtained by solving matching conditions between the boundary layer and the main flow in the guide (section 2-1 of \cite{Sugimoto91}).

The Riemann-Liouville fractional integral of order 1/2 of a causal function $w(t)$ (thus $w(0)=0$) is defined by
\begin{equation}
\frac{\textstyle \partial^{-1/2}}{\textstyle \partial t^{-1/2}}w(t)=\frac{\textstyle H(t)}{\textstyle \sqrt{\pi\,t}}*w=\frac{\textstyle 1}{\textstyle \sqrt{\pi}}\int_0^t(t-\tau)^{-1/2}w(\tau)\,d\tau,
\label{RiemannLiouville}
\end{equation}
where * is the convolution product in time, and $H(t)$ is the Heaviside step function \cite{Podlubny99}.


\subsubsection{Determination of the flux $F^\pm$}

\begin{figure}[htbp]
\begin{center}
\begin{tabular}{cc}
(a) & (b)\\
\hspace{-1cm}
\includegraphics[scale=0.65]{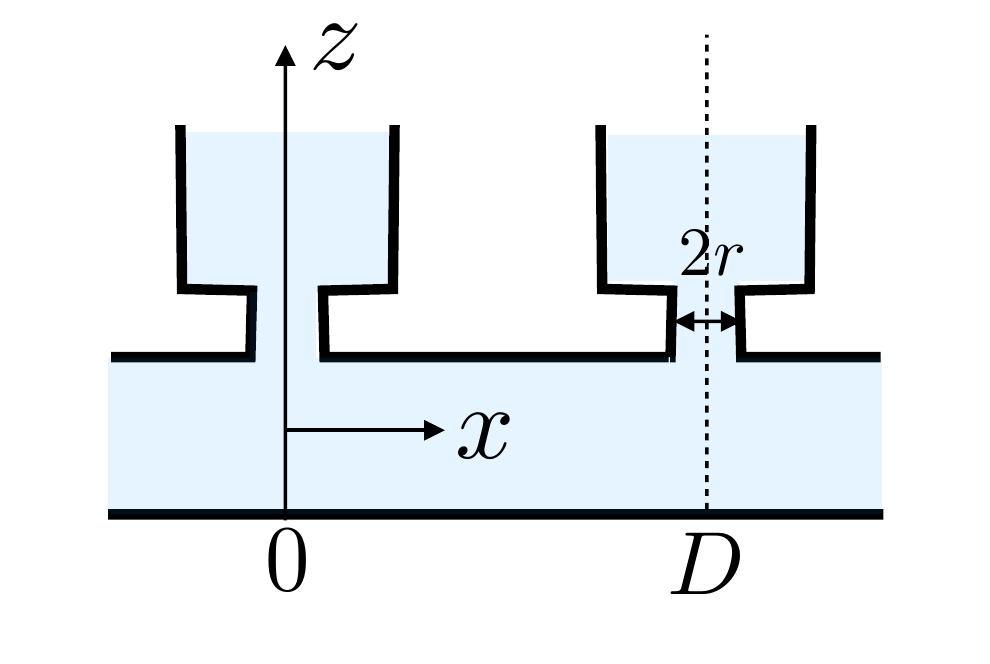}
 &
\includegraphics[scale=0.65]{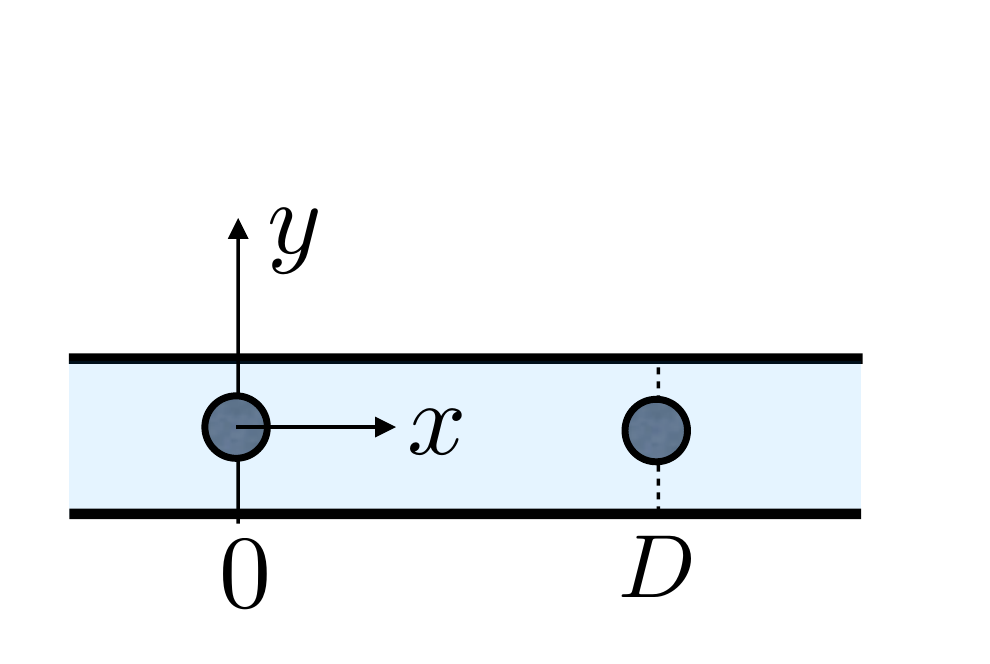}
\end{tabular}
\caption{\label{vues} portion of length $D$ of the tube. View from the side (a). View from above (b).}
\end{center}
\end{figure}

For convenience, $\Sigma$ is taken as the part $x \in ]-r,D-r[$, the part $\Sigma_B$ being then centered at $x=0$, located in $x \in ]-r,r[$ (see Fig. \ref{vues}). Then the flux $F^\pm$ can be determined at each position $x$, depending on two cases.

\begin{figure}[htbp]
\begin{center}
\begin{tabular}{cc}
(a) & (b)\\
\hspace{-1cm}
\includegraphics[scale=0.50]{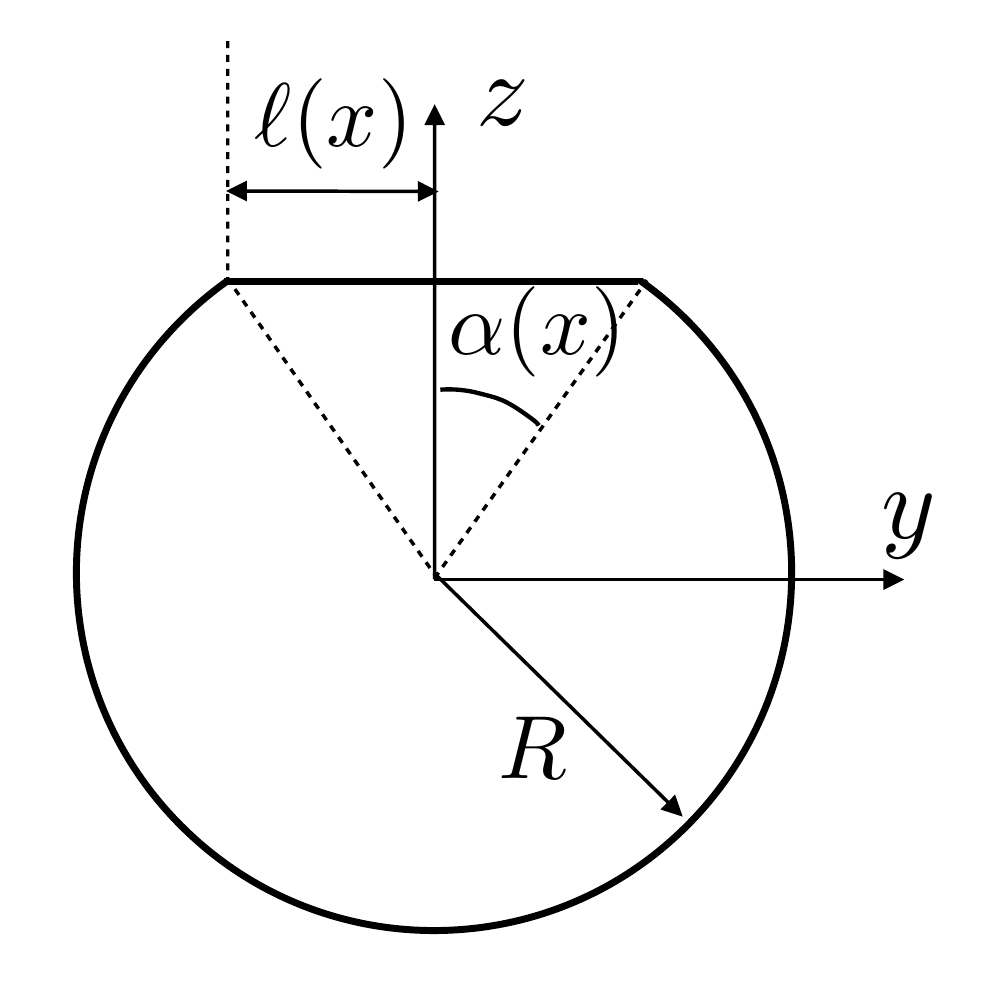}
 &
\hspace{0.5cm}
\includegraphics[scale=0.55]{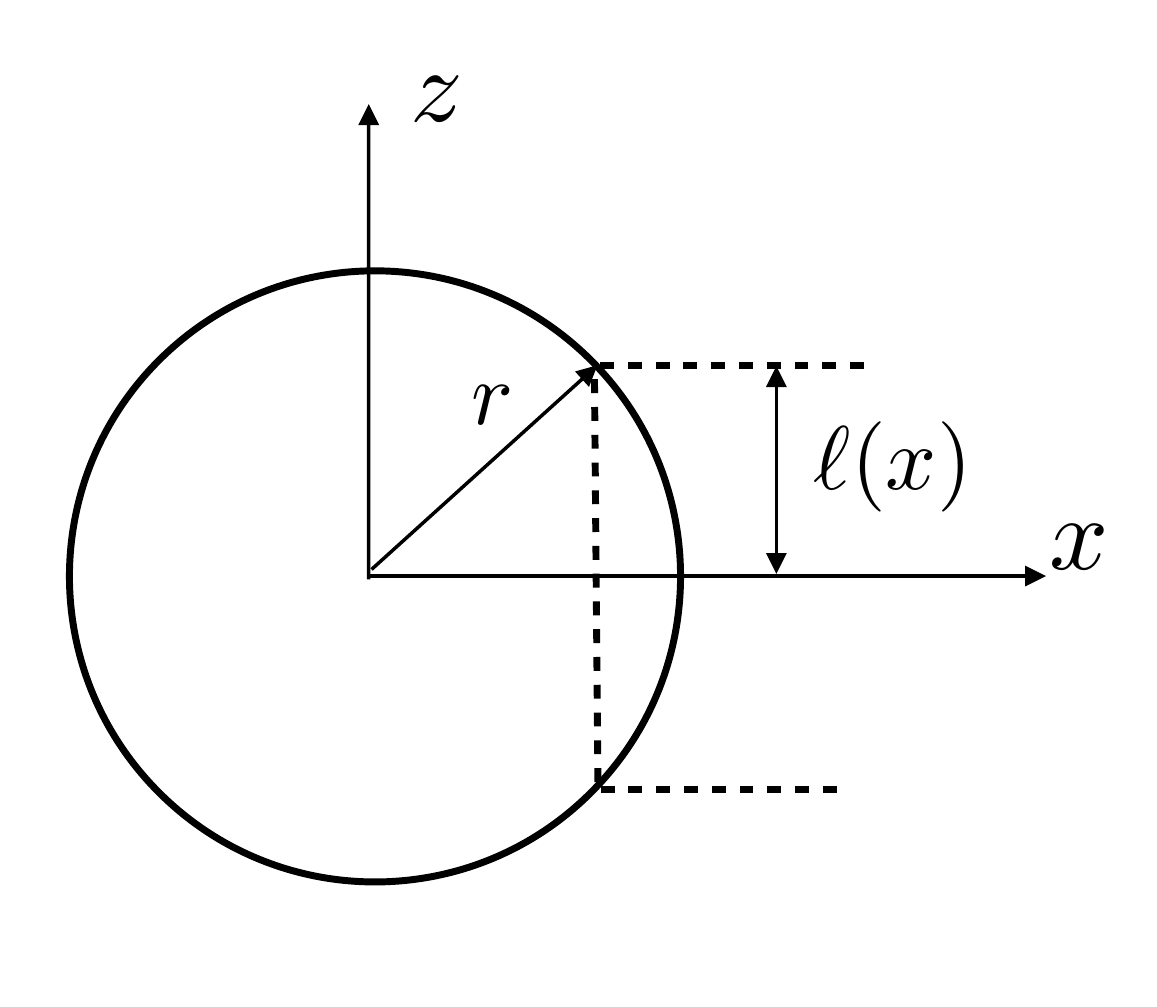}
\end{tabular}
\caption{\label{coupe} (a) $\sigma(x)$ for $x \in ]-r,r[$; (b) view from above of the resonators in $x \in ]-r,r[$.}
\end{center}
\end{figure}

\noindent \underline{Case 1:} $x \in ]r,D-r[$. Then $\sigma(x)$ is a disc of radius $R$ and $\rho v_n^\pm=\rho_0 v_b^\pm$ on $\sigma$. Therefore
\begin{eqnarray*}
F^\pm(x) &=& \frac{1}{A} \int_0^{2 \pi} (\rho v_n^\pm)(x,R \sin \theta, R \cos \theta) R d\theta,\\
&=& \displaystyle \frac{1}{A} \left(2 \pi R \right) \rho_0 v_b^\pm(x),
\end{eqnarray*}
where $\theta$ is measured with respect to the $z$ axis (as $\alpha$ in Fig. \ref{coupe}(a)).\\

\noindent \underline{Case 2:} $x \in ]-r,r[$. A cross section of the tube at $x \in ]-r,r[$ is sketched in Fig. \ref{coupe}(a). The integration path $\sigma(x)$ is made of two parts: (i) a line of length $2 \ell(x)$ on which $\rho v_n^\pm = - Q$; (ii) a portion of a circle of radius $R$, between the angles $\theta = \alpha$ and $\theta = 2 \pi - \alpha$, on which $\rho v_n^\pm = \rho_0 v_b^\pm$. Using Fig. \ref{coupe}(b) representing $\Sigma_B$ seen from above, it follows that:
\begin{equation} \label{}
\ell(x) = \left\{ \begin{array}{rcl}
 0 & \mbox{for} & x \in ]r,D-r[,\\[8pt]
 \sqrt{r^2-x^2} & \mbox{for} & x \in ]-r,r[.
\end{array} \right. \end{equation}
Using Fig. \ref{coupe}(a), one finds also that $\alpha(x) = \ds \arcsin \left( \ell(x)/R \right)$. Then (\ref{F}) becomes:
\begin{eqnarray*}
F^\pm(x) &=& \frac{1}{A} \left[ \int_{\alpha(x)}^{2 \pi-\alpha(x)} (\rho_0 v_b^\pm)(x,R \sin \theta, R \cos \theta) R d\theta
- \int_{-\ell(x)}^{\ell(x)} Q(x) ds \right],\\
&=& \displaystyle \frac{2}{A} \left\{ \left[\pi - \alpha(x) \right] R \rho_0 v_b^\pm(x) - Q(x) \ell(x) \right\}.
\end{eqnarray*}

To sum up, using the values of $Q$ and $v_b^\pm$ from (\ref{B}), one deduces that for all $x \in ]-r,D-r[$:
\begin{equation} 
\label{flux} 
\frac{a_0}{2 \rho_0} F^\pm = c(x) \frac{\textstyle \partial^{-1/2}}{\textstyle \partial t^{-1/2}}\frac{\textstyle \partial u}{\textstyle \partial x}^\pm e(x) (1-2 m p) \frac{\textstyle \partial p}{\textstyle \partial t},
\end{equation}
where we have introduced the functions
\begin{equation} 
\label{funccete} 
\left\{ 
\begin{array}{rcl}
c(x) &=& c_0 \displaystyle \left[ 1 - \frac{1}{\pi} \arcsin \left( \frac{\ell(x)}{R} \right) \right],\\
e(x) &=& e_0 \ell(x),
\end{array} 
\right. 
\end{equation}
with the coefficients
\begin{equation} \label{coeffbis}
\begin{array}{l}
\displaystyle
c_0=\frac{\textstyle C \, a_0 \sqrt{\nu}}{\textstyle R},\quad e_0 = \frac{\textstyle V}{\textstyle \rho_0\,a_0\,A B},
\end{array}
\end{equation}
and $V$ the volume of the resonator at $x=0$.

This result has been obtained for a unit part of the tube located in $x \in ]-r,D-r[$. 
\begin{figure}[htbp]
\begin{center}
\includegraphics[width=12.cm,clip]{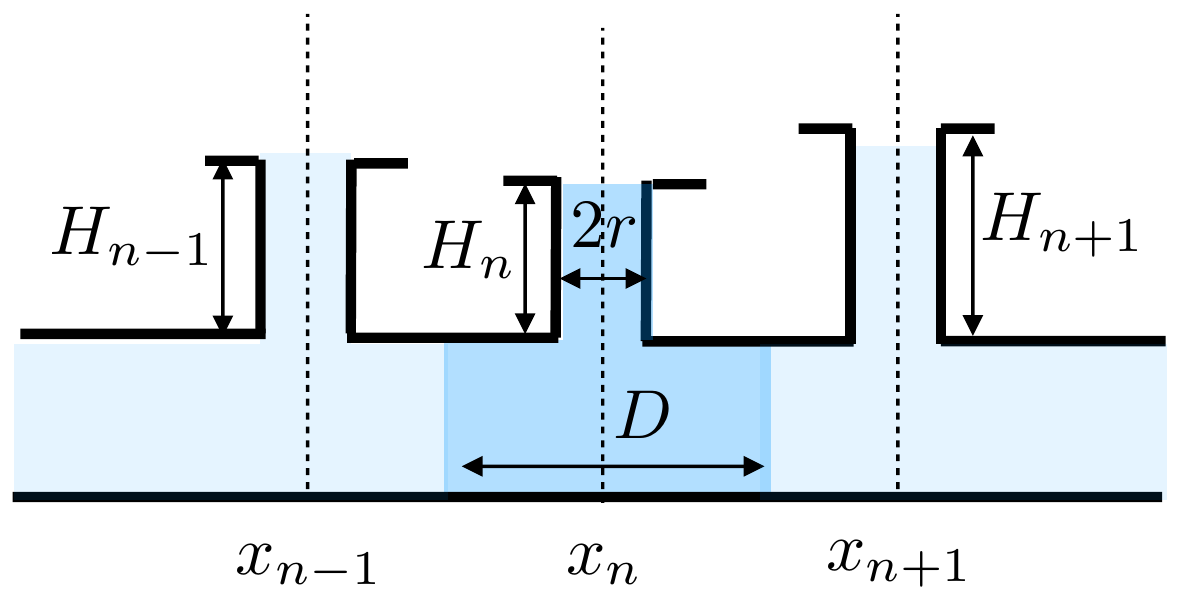}
\caption{\label{resonateurn} side view of the guide connected to Helmholtz resonators. The $i$-th cell is highlighted in grey.}
\end{center}
\end{figure}
Now, we consider the entire tube, in which the resonators are centered at $x_i=i D$, $i=1,2, \cdots$ (see Fig. \ref{resonateurn}). We note $V_i$ the volume (and $H_i$ the height) of the $i$-th resonator at $x_i$. In the $i$-th cell $x \in [x_i-D/2, x_i+D/2]$ (represented in grey on Fig. \ref{resonateurn}) and based on (\ref{flux}), Eq. (\ref{invariant}) takes the form
\begin{equation} \label{varu}
\ds\frac{\partial u}{\partial t}^\pm + \frac{\partial}{\partial x} \left(\pm a_0 u^\pm +b\ds \frac{(u^\pm)^2}{2}\right) \ds \mp c(x) \frac{\partial^{-1/2}}{\partial t^{-1/2}} \frac{\partial u}{\partial x} ^\pm - d \frac{\partial^2 u}{\partial x^2}^\pm = \mp e(x) (1-2 m p) \ds \frac{\partial p}{\partial t}.
\end{equation}
The coefficients $c(x)$ and $e(x)$ are defined by
\begin{equation} \label{cetedex} 
\left\{ 
\begin{array}{lcl}
c(x) &=& \ds c_0 \left[1 - \frac{1}{\pi} \arcsin \left( \frac{\ell_i(x)}{R} \right) \right],\\
e(x) &=& e_i \ell_i(x),
\end{array} 
\right.
\end{equation}
where
\begin{equation} \label{ln} 
\ell_i(x) = 
\left\{ 
\begin{array}{lcl}
0 & \mbox{for} & r < |x-x_i| \leq \ds \frac{D}{2},\\
\sqrt{r^2-(x-x_i)^2} & \mbox{for} & |x-x_i| \leq r,
\end{array} 
\right.
\end{equation}
$c_0$ is defined in (\ref{coeffbis}), and 
\begin{equation} \label{en}
\begin{array}{l}
\ds
e_i = \frac{V_i}{\rho_0\,a_0\,A B}.
\end{array}
\end{equation}

\subsubsection{Reformulation of the viscothermal losses}

The fractional integrals in (\ref{varu}) have two drawbacks: they prevent from obtaining an energy balance, and the dispersion relation is singular at zero frequency \cite{Lombard16}. It is therefore preferable to replace them by fractional derivatives. Based on the linear approximation $(\partial u^\pm/\partial t) \pm a_0(\partial u^\pm/\partial x) \approx 0$, one has
\begin{equation}
\frac{\partial^{-1/2}}{\partial t^{-1/2}}\frac{\partial u^\pm}{\partial x}\approx \mp \frac{1}{a_0}\frac{\partial^{1/2}}{\partial t^{1/2}}u^\pm.
\label{Dm12}
\end{equation}
The rigorous derivation of this step is detailed in \ref{SecFractional}. Injecting (\ref{Dm12}) in (\ref{varu}) yields finally the first two equations of (\ref{NewModel}). 


\subsection{Equation in the resonators}
 
The excess pressure $p$ in a resonator of volume $V$ satisfies the equation \cite{Monkewitz85a,Monkewitz85b}:
\begin{equation} \label{resonator}
\frac{\textstyle \partial^2 p}{\textstyle \partial t^2}+f\frac{\textstyle \partial^{3/2} p}{\textstyle \partial t^{3/2}} + g p-m\frac{\textstyle \partial^2 p^2}{\textstyle \partial t^2}+n\left|\frac{\textstyle \partial p}{\textstyle \partial t}\right|\,\frac{\textstyle \partial p}{\textstyle \partial t} = g p',
\end{equation}
where
\begin{equation} \label{coefres}
f=\frac{\textstyle 2\,\sqrt{\nu}}{\textstyle r}\,\frac{\textstyle L^{'}}{\textstyle L_e},\quad g=\frac{a_0^2 \textstyle B}{\textstyle L_e\,V},\quad n=\frac{\textstyle V}{\textstyle B\,L_e\,\rho_0\,a_0^2},
\end{equation}
and where $p'$ is the excess pressure at the mouth of the tube. The coefficients $m$ in (\ref{m}) and $n$ in (\ref{coefres}) describe nonlinear processes in the resonators. The semi-empirical coefficient $n$ accounts for the jet loss resulting from the difference in inflow and outflow patterns \cite{Sugimoto92,Sugimoto04}. These nonlinear processes have to be included to get a good agreement with the experimental measurements \cite{Richoux15}. The Caputo fractional derivative of order 3/2 is obtained by applying (\ref{RiemannLiouville}) to $\partial^2 p/\partial t^2$. Corrected lengths have been introduced: $L^{'}=L+2\,r$ accounts for the viscous end corrections, and the corrected length $L_e=L+\eta$ accounts for the end corrections at both ends of the neck, where $\eta\approx 0.82\,r$ is determined experimentally \cite{Sugimoto92}.

In the one-way model \cite{Sugimoto92}, a linear approximation is used to link the right-going fields in the tube: $p'^+ = (\gamma p_0/a_0) u^+$. In a symmetric way, $p'^- = -(\gamma p_0/a_0) u^-$. Here $p'$ is induced by both the right-going waves and the left-going waves; assuming linearity gives
\begin{equation} \label{pprime}
p' = p'^+ + p'^- = \frac{\gamma p_0}{a_0} (u^+ - u^-).
\end{equation}
Injecting (\ref{pprime}) in (\ref{resonator}) leads to
\begin{equation} \label{sugip2way}
\frac{\partial^2 p}{\partial t^2}+f\frac{\partial^{3/2} p}{\partial t^{3/2}} + g p-m\frac{\partial^2 p^2}{\partial t^2}+n\left|\frac{\partial p}{\partial t}\right|\,\frac{\partial p}{\partial t} = h (u^+ - u^-),
\end{equation}
with $h=g (\gamma\,p_0/ a_0)$.

For resonators of variable volume, Eq. (\ref{sugip2way}) becomes, for $|x-x_i| \leq r$ ($i=1, 2,\cdots$)
\begin{equation}\label{varp}
\ds\frac{\partial^2 p}{\partial t^2} + f \frac{\partial^{3/2} p}{\partial t^{3/2}} + g(x) p-m\frac{\partial^2 p^2}{\partial t^2} + n(x) \left|\frac{\partial p}{\partial t}\right|\,\frac{\partial p}{\partial t} = h(x) (u^+ - u^-),
\end{equation}
with
\begin{equation} \label{getn}
g(x) = \ds \frac{a_0^2 B}{L_e\,V_i},
\quad
n(x) = \ds \frac{V_i}{B\,L_e\,\rho_0\,a_0^2},
\quad
h(x) = \frac{\gamma\,p_0}{a_0} g(x).
\end{equation}
One recognizes the third equation in (\ref{NewModel}). The functions $g(x)$, $h(x)$ and $n(x)$ depend on $x$ because the resonators have different heights, but they are constant on each resonator. They just depend on $i$, the index of the resonators. The equation (\ref{varp}) is not solved outside the resonators (that is on $r < |x-x_i| \leq \ds D/2$) contrary to Sugimoto's model recalled in (\ref{sugip}).

In \ref{SecSugi}, we show that the original model of Sugimoto \cite{Sugimoto92} can be recovered from the new model (\ref{NewModel}). Note also that (\ref{NewModel}) can be easily extended to the case of resonators not periodically placed; only the values of $x_i$ need to be changed.


\section{First-order system}\label{SecEDP}

In this part, Eq. (\ref{NewModel}) is formulated as a first-order system, which enables us to determine an energy balance and to build a numerical scheme.

\subsection{Diffusive approximation}\label{SecEDPDF}

The fractional derivatives in (\ref{NewModel}) are nonlocal in time and they rely on the full history of the solution, which is numerically memory-consuming. An alternative approach is based on a diffusive representation of fractional derivatives, and then on its approximation. This method has already been presented in \cite{Richoux15,Lombard14} and we just recall it briefly. The half-order integral (\ref{RiemannLiouville}) of $w(t)$ is written
\begin{equation}
\frac{\partial^{-1/2}}{\partial t^{-1/2}}w(t) \simeq \sum_{\ell=1}^{N}\mu_{\ell}\,\varphi_{\ell}(t),
\label{RDI12}
\end{equation}
where the diffusive variables $\varphi_{\ell}(t)=\varphi(t,\theta_{\ell})$ satisfy the ODE
\begin{equation}
\left\{
\begin{array}{l}
\ds
\frac{\partial \varphi_{\ell}}{\partial t}=-\theta_{\ell}^2\,\varphi_{\ell}+\frac{\textstyle 2}{\textstyle \pi}\,w,\\
[8pt]
\varphi_{\ell}(0)=0.
\end{array}
\right.
\label{ODEI12}
\end{equation}
The approximation (\ref{RDI12}) follows from the approximation of an integral thanks to a quadrature formula on $N$ points, with weights $\mu_{\ell}$ and nodes $\theta_{\ell}$, which are issued from an optimization process.

To get fractional derivatives of orders $1/2$ and $3/2$, we differentiate (\ref{RDI12}) in terms of $t$. Then we deduce:
\begin{equation}
\frac{\partial^{1/2} w}{\partial t^{1/2}} = \frac{\partial}{\partial t} \frac{\partial^{-1/2} w}{\partial t^{-1/2}} \simeq \sum_{\ell=1}^{N} \mu_{\ell}\, \frac{\partial \varphi_{\ell}}{\partial t}(t) = \mu_{\ell} \left( -\theta_{\ell}^2\,\varphi_{\ell} + \frac{2}{\pi}\,w \right).
\label{RDD12}
\end{equation}
Similarly, the derivative of order 3/2 is written
\begin{equation}
\frac{\partial^{3/2} w}{\partial t^{3/2}} = \frac{\partial}{\partial t} \frac{\partial^{1/2} w}{\partial t^{1/2}} \simeq \sum_{\ell=1}^{N} \mu_{\ell}\, \frac{\partial \xi_{\ell}}{\partial t}(t),
\label{RDD32}
\end{equation}
where we have introduced $\xi_{\ell} = \partial \varphi_{\ell}/\partial t$.
The diffusive variable $\xi_{\ell}(t)$ satisfies the following ODE (derivative of (\ref{ODEI12})):
\begin{equation}
\left\{
\begin{array}{l}
\ds
\frac{\partial \xi_{\ell}}{\partial t}=-\theta_{\ell}^2 \,\xi_{\ell}+\frac{\textstyle 2}{\textstyle \pi}\,\frac{\partial w}{\partial t},\\
[8pt]
\xi_{\ell}(0)=0.
\end{array}
\right.
\label{ODED12}
\end{equation}
The initial condition is obtained using $w(0)=0$, since we consider causal data. Thanks to these diffusive approximations, Eq. (\ref{NewModel}) can be written as the following system 
\begin{subnumcases}{\label{EDPbis}}
\ds \nonumber
\frac{\partial u}{\partial t}^\pm+\frac{\partial}{\partial x}\left(\pm a_0 u^\pm + b \frac{(u^\pm)^2}{2}\right)\\
= - \frac{c}{a_0} \sum_{\ell=1}^N\mu_{\ell}\left( -\theta_{\ell}^2 \varphi^\pm_{\ell}+\frac{2}{\pi} u^\pm \right) +d\frac{\partial^2 u}{\partial x^2}^\pm \mp e (1 - 2 m p) q,\label{EDP1bis}\\
\ds
\frac{ \partial p}{\partial t}=q,\label{EDP2bis}\\
\ds
\frac{\partial q}{\partial t}=h(u^+-u^-)-g p-f\sum_{\ell=1}^N\mu_{\ell} \left( -\theta_{\ell}^2\xi_{\ell}+\frac{2}{\pi}q \right) +m\frac{ \partial^2 p^2}{\partial t^2}-n\left|q\right|q,\label{EDP3bis}\\
\ds
\frac{\partial \varphi^\pm_{\ell}}{\partial t}=-\theta_{\ell}^2 \varphi^\pm_{\ell}+\frac{2}{\pi} u^\pm,\hspace{1.5cm} \ell=1\cdots N,\label{EDP5bis}\\
[6pt]
\ds
\frac{\partial \xi_{\ell}}{\partial t}=-\theta_{\ell}^2 \xi_{\ell}+\frac{2}{\pi}q,\hspace{2.0cm} \ell=1\cdots N.\label{EDP6bis}
\end{subnumcases}
Contrary to \cite{Lombard14}, all the initial conditions are null: $u^\pm=0$, $p=0$, $\frac{\partial p}{\partial t}=0$. If this were not the case, then the diffusive representation of the 3/2 derivative would imply non-null initial conditions on $\xi_{\ell}$ in (\ref{ODED12}). The interested reader is referred to \cite{Lombard16} for additional details on this topic.


\subsection{Energy balance}

The system (\ref{EDPbis}) is suitable to define an energy and to prove the energy decreasing, in an infinite medium and for smooth solutions:

\begin{resultat} \label{energie}
Let
\begin{equation}
\displaystyle
{\cal E}={\cal E}_1^++{\cal E}_1^-+{\cal E}_2 \; \mbox{ and } \; {\cal K}={\cal K}_1^++{\cal K}_1^-+{\cal K}_2,
\label{EK}
\end{equation}
with
\begin{subnumcases}{\label{NRJ}}
\ds
{\cal E}_1^\pm =\frac{1}{2}\int_{\mathbb{R}}\left((u^\pm)^2+\frac{\pi}{2}\,\frac{c}{a_0} \sum_{\ell=1}^N\mu_{\ell}\,\theta_{\ell}^2\,(\varphi_{\ell}^\pm)^2 \right)\,dx,\label{E1}\\
[6pt]
\ds
{\cal E}_2=\frac{1}{2}\int_\mathbb{R}\left(\frac{eg}{h}p^2+\frac{e}{h}(1-2\,m\,p)\,q^2+\frac{\pi}{2}\frac{ef}{h}\sum_{\ell=1}^N\mu_{\ell}\,\theta_{\ell}^2\,\xi_{\ell}^2\right)\,dx,\label{E2}\\
[6pt]
\ds
{\cal K}_1^\pm =\frac{\pi}{2}\int_\mathbb{R} \frac{c}{a_0} \sum_{\ell=1}^N\mu_\ell\left(\frac{\partial \varphi_\ell^\pm}{\partial t}\right)^2\,dx+\int_\mathbb{R}d\left(\frac{\partial u^\pm}{\partial x}\right)^2\,dx,\\
[6pt]
\ds
{\cal K}_2=\int_\mathbb{R}\left(\frac{e\,n}{h}q^2\left(|q|-\frac{m}{n}q\right)+\frac{\pi}{2}\frac{e\,f}{h}\sum_{\ell=1}^N\mu_{\ell}\left(\frac{\partial \xi_{\ell}}{\partial t}\right)^2\,dx\right).
\end{subnumcases}
We neglect the $2 m p$ terms in the tube, ie in the evolution equations of $u^{\pm}$ in (\ref{NewModel}). Then the following energy balance holds:
\begin{equation}
\frac{d{\cal E}}{dt}=-{\cal K}.
\label{dEdt}
\end{equation}
\end{resultat}

The proof is reported in \ref{SecNRJproof}. Four remarks are raised by (\ref{dEdt}).
\begin{itemize}
\item ${\cal E}_1^\pm$ and ${\cal K}_1^\pm$ involve quantities in the tube, notably the kinetic energy $(u^\pm)^2$; ${\cal E}_2$ and ${\cal K}_2$ involve quantities in the resonators, notably the potential energy with terms proportional to $p^2$ and $q^2$. 
\item We have not succeeded in obtaining a proof by accounting for the $2 m p$ term in the advection equations. Nevertheless, the hypothesis $2 m p \simeq 0$ is consistent with Sugimoto's work, where only the influence of $m$ in the resonators equation has been considered: see (\ref{sugiu}). It is also consistent with the hypothesis of a weak nonlinear regime. Indeed, as shown in \cite{Richoux15,Lombard14}:
\begin{equation}
2 m p \approx \displaystyle (\gamma-1)\,\frac{u^+ - u^-}{a_0},
\end{equation}
which is lower than $1$ under the hypothesis of weak nonlinearity $|u^\pm| \ll a_0$. 
\item The term ${\cal E}$ in (\ref{EK}) is positive if $\mu_\ell >0$ and $1-2mp>0$. The first condition is imposed when the coefficients of the diffusive representation are determined during the optimisation process \cite{Richoux15}. The second condition is satisfied in the weak nonlinear regime.
\item The term ${\cal K}$ in (\ref{EK}) is positive if $\mu_\ell >0$ and $m<n$. The first condition has been already discussed. The second condition reads $m/n=\frac{\gamma-1}{2}\,\frac{B L_e}{V}<1$, where $B L_e$ is the resonator neck volume, and $V$ is the volume of the resonators. For the experimental configuration under study, this ratio is lower than 1 and thus the condition is satisfied.
\end{itemize}


\subsection{Numerical scheme}\label{SecEDPSys}

In this part, we present the numerical resolution of (\ref{EDPbis}) with null initial conditions. Source terms $s^\pm(t)$ at $x_{s^\pm}$ model the acoustic sources of right-going waves and left-going waves: $u^+(x_{s^+},t) = s^+(t)$, $u^-(x_{s^-},t) = s^-(t)$. For numerical purpose, it is necessary to write the evolution equations as a first-order system in time. To do so, the term $\partial^2 p^2/\partial t^2$ in Eq. (\ref{EDP3bis}) is expanded and leads to
\[
(1 - 2 m p) \frac{\partial q}{\partial t} = h(u^+ - u^-)-g\,p-f\,\sum_{\ell=1}^N\mu_{\ell} \left( -\theta_{\ell}^2\,\xi_{\ell}+\frac{2}{\textstyle \pi}\,q \right) + 2 m q^2 - n \left|q\right| q.
\]
The $(4+3\,N)$ unknowns for the counter-propagating waves are gathered in the vector
\begin{equation}
{\bf U}=\left(u^+,u^-,\,p,\,q,\varphi^+_1,\cdots,\,\varphi^+_N,\varphi^-_1,\cdots,\,\varphi^-_N,\,\xi_1,\cdots,\,\xi_N\right)^T.
\label{VecU}
\end{equation}
Then the nonlinear systems (\ref{EDPbis}) is written in the form
\begin{equation}
\frac{\textstyle \partial}{\textstyle \partial t}{\bf U}+\frac{\textstyle \partial}{\textstyle \partial x}{\bf F}({\bf U})={\bf G}\,\frac{\textstyle \partial^2}{\textstyle \partial x^2}{\bf U}+{\bf S}({\bf U}),
\label{SystHyper}
\end{equation}
where ${\bf F}$ is the flux function
\begin{equation}
{\bf F}=\left(a_0 u^+ + b \,\frac{\textstyle (u^+)^2}{\textstyle 2}, - a_0 u^- + b \, \frac{\textstyle (u^-)^2}{\textstyle 2},\,0,\,0,\cdots,\,0\right)^T,
\label{Fnonlin}
\end{equation}
and where the source term ${\bf S}$ is
\begin{equation}
{\bf S}=
\left(
\begin{array}{l}
\ds
- \frac{c}{a_0} \sum_{\ell=1}^N \mu_{\ell} \left( -\theta_{\ell}^2 \, \varphi^+_{\ell}+\frac{2}{\pi}\,u^+ \right) - e (1 - 2 m p) \, q\\
[10pt]
\ds
- \frac{c}{a_0} \sum_{\ell=1}^N \mu_{\ell} \left( -\theta_{\ell}^2 \, \varphi^-_{\ell}+\frac{2}{\pi}\,u^- \right) + e (1 - 2 m p) \, q\\
[10pt]
q\\
[6pt]
\displaystyle
\frac{\textstyle 1}{\textstyle 1-2 m p}\left( h(u^+-u^-)-g\,p-f\,\sum_{\ell=1}^N\mu_{\ell} \left( -\theta_{\ell}^2\,\xi_{\ell}+\frac{2}{\textstyle \pi}\,q \right) + 2 m q^2 - n \left|q\right| q \right)\\
\\
\ds
-\theta_{\ell}^2\,\varphi^+_{\ell}+\frac{2}{\pi}\,u^+,\hspace{1cm} \ell=1\cdots N\\
\\
\ds
-\theta_{\ell}^2\,\varphi^-_{\ell}+\frac{2}{\pi}\,u^-,\hspace{1cm} \ell=1\cdots N\\
\\
\ds
-\theta_{\ell}^2\,\xi_{\ell}+\frac{\textstyle 2}{\textstyle \pi} \, q,\hspace{1.5cm} \ell=1\cdots N
\end{array}
\right).
\label{Snonlin}
\end{equation}
As soon as $m \neq 0$ and $n \neq 0$, ${\bf S}({\bf U})$ is no longer a linear operator ($m=0=n$ has been considered in \cite{Lombard14}). The Jacobian matrix $\frac{\partial {\bf F}}{\partial {\bf U}}$ in (\ref{Fnonlin}) is diagonalizable with  real eigenvalues: $a_0+b\,u^+$, $-a_0+b\,u^-$ and 0 with multiplicity $3\,N+2$, which ensures propagation with finite velocity. These eigenvalues do not depend on the quadrature coefficients $\mu_{\ell}$ and $\theta_{\ell}$. The  diagonal matrix ${\bf G}=\mbox{diag}(d,d,\,0,\cdots,\,0)$ incorporates the volume attenuation.

Let us describe briefly the numerical procedure; additional details can be found in \cite{Richoux15,Lombard14}. To compute the $N$ quadrature coefficients $\mu_{\ell}$ and $\theta_{\ell}$ in (\ref{Snonlin}), we use a nonlinear optimization with the positivity constraints $\mu_{\ell}\geq 0$ and $\theta_{\ell}\geq 0$ \cite{Richoux15}. In order to integrate the system (\ref{SystHyper}), a grid is introduced, with a uniform spatial mesh size $\Delta x$ and a variable time step $\Delta t_n$. The computations are done with $N_x=1000$ grid nodes. The approximation of the exact solution ${\bf U}(x_j = j\,\Delta\,x, t_n = t_{n-1}+\,\Delta t_n)$ is denoted by ${\bf U}_j^n$. A stability analysis \cite{Lombard14} leads to the restriction on the time step
\begin{equation}
\beta= \frac{a_{\max} ^{(n)} \Delta t_n}{\Delta x} \left(1+\frac{\textstyle 1}{\textstyle \mbox{Pe}}\right) \leq 1.
\label{Dt}
\end{equation}
In the CFL condition (\ref{Dt}), $\mbox{Pe}=a_{\max} ^{(n)}\,\Delta x/2\,d\approx10^5$ is the discrete P\'eclet number, with the maximal velocity
\begin{equation}
a_{\max} ^{(n)}=a_0+b\,\max_j [ \max(|u_j^{n+}|,|u_j^{n-}|)].
\end{equation}
In practice, we choose $\beta=0.95$. To treat the system (\ref{SystHyper}), a Strang splitting is used \cite{Toro99}, ensuring both simplicity and efficiency: the original equation (\ref{EDPbis}) is split in a propagative equation and a forcing equation, which are solved successively with adequate time increments. The propagative part is solved by a standard second-order TVD scheme (a finite-volume scheme with flux limiters) for nonlinear hyperbolic PDE \cite{LeVeque92} combined with a centered finite-difference approximation \cite{Lombard14}.


\section{Results}\label{SecResults}

\begin{table}[htbp]
\begin{center}
\begin{small}
\begin{tabular}{|l|l|l|l|l|l|}
\hline
$\gamma$ & $p_0$ (Pa) & $\rho_0$ (kg/m$^{3}$) & $Pr$    & $\nu$ (m$^2$/s) & $\mu_v/\mu$\\
\hline  
1.403    & $ 10^5$    & 1.177                 & 0.708   & $1.57\,10^{-5}$ & 0.60       \\
\hline
\hline
$R$ (m)  & $D$ (m)    & $r$ (m)               & $L$ (m) & $r_h$ (m)       & $H$ (m)    \\
\hline
0.025    & 0.1        & 0.01                  & 0.02    & 0.0215          & 0.02, 0.07 or 0.13 \\
\hline
\end{tabular}
\end{small} 
\end{center}
\vspace{-0.5cm}
\caption{Physical parameters of the air at $15\,^{\circ}\mathrm{C}$, and geometrical parameters of the tube with resonators.}
\label{TabParam}
\end{table}

\subsection{Test 1: comparison with experiments}\label{SecTest1}

The first result is based on the experimental measurements detailed in \cite{Richoux15}. The reader is referred to this paper for a complete description of the experimental and numerical setups. The physical and geometrical parameters are given in table \ref{TabParam}. The explosion of a balloon at $x=0$ generates a shock wave, which propagates along a tube of length $6.15$ m. This wave interacts with resonators regularly spaced from $x=0.2$ m up to the end of the tube. Two models are compared: the one-way model with constant coefficients (ie the original Sugimoto's model used in \cite{Richoux15}) and the two-way model with variable coefficients developed here. The computations are done on 5000 grid nodes. A preliminary numerical study has been performed to verify that this discretization is sufficiently fine to capture the converged solution, at the scale of the figures. The same remark holds for all the forthcoming tests.

\begin{figure}[h!]
\begin{center}
\begin{tabular}{ccc}
(a) & (b)\\
\hspace{-1.2cm}
\includegraphics[scale=0.36]{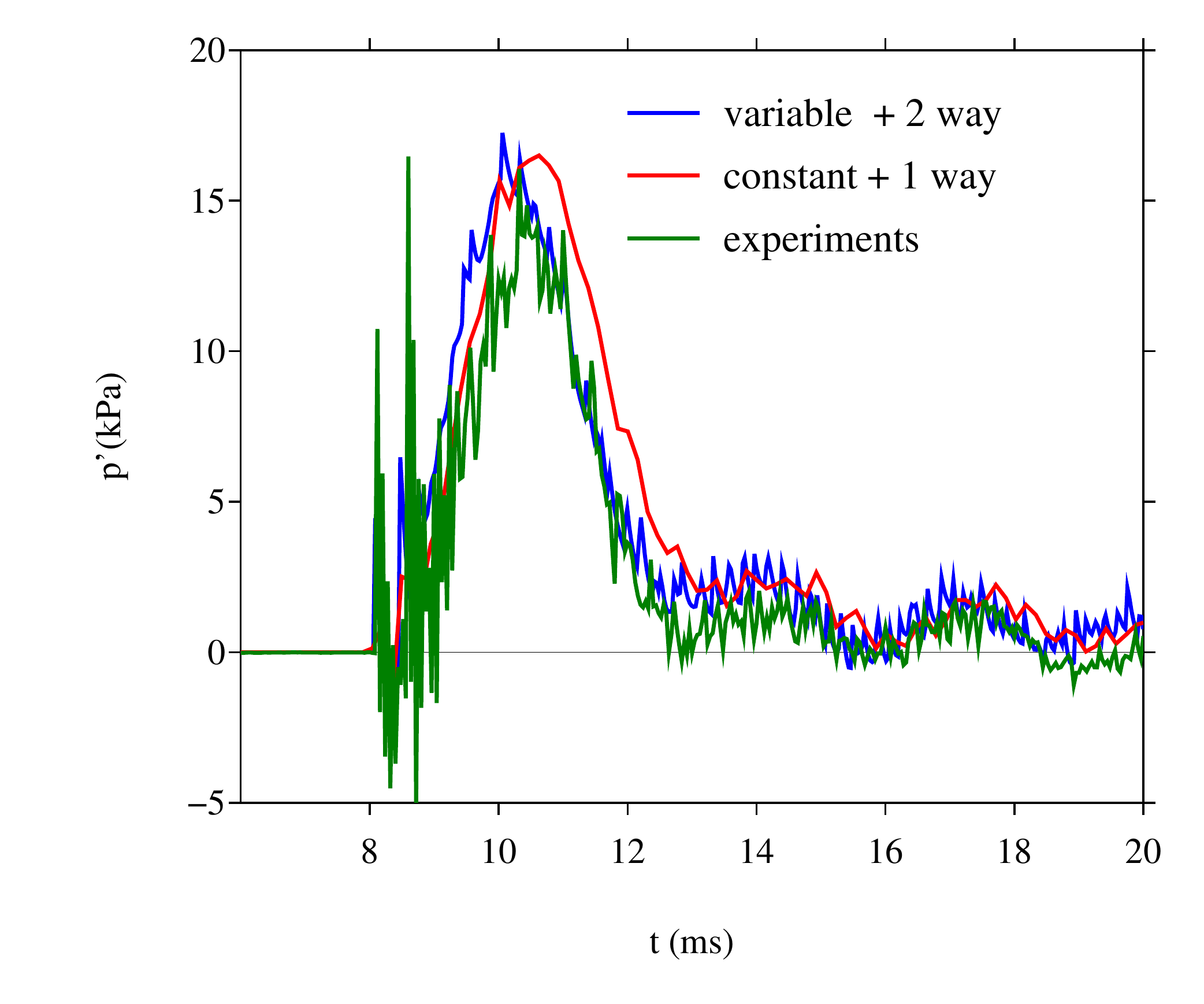} &
\hspace{-1.2cm}
\includegraphics[scale=0.36]{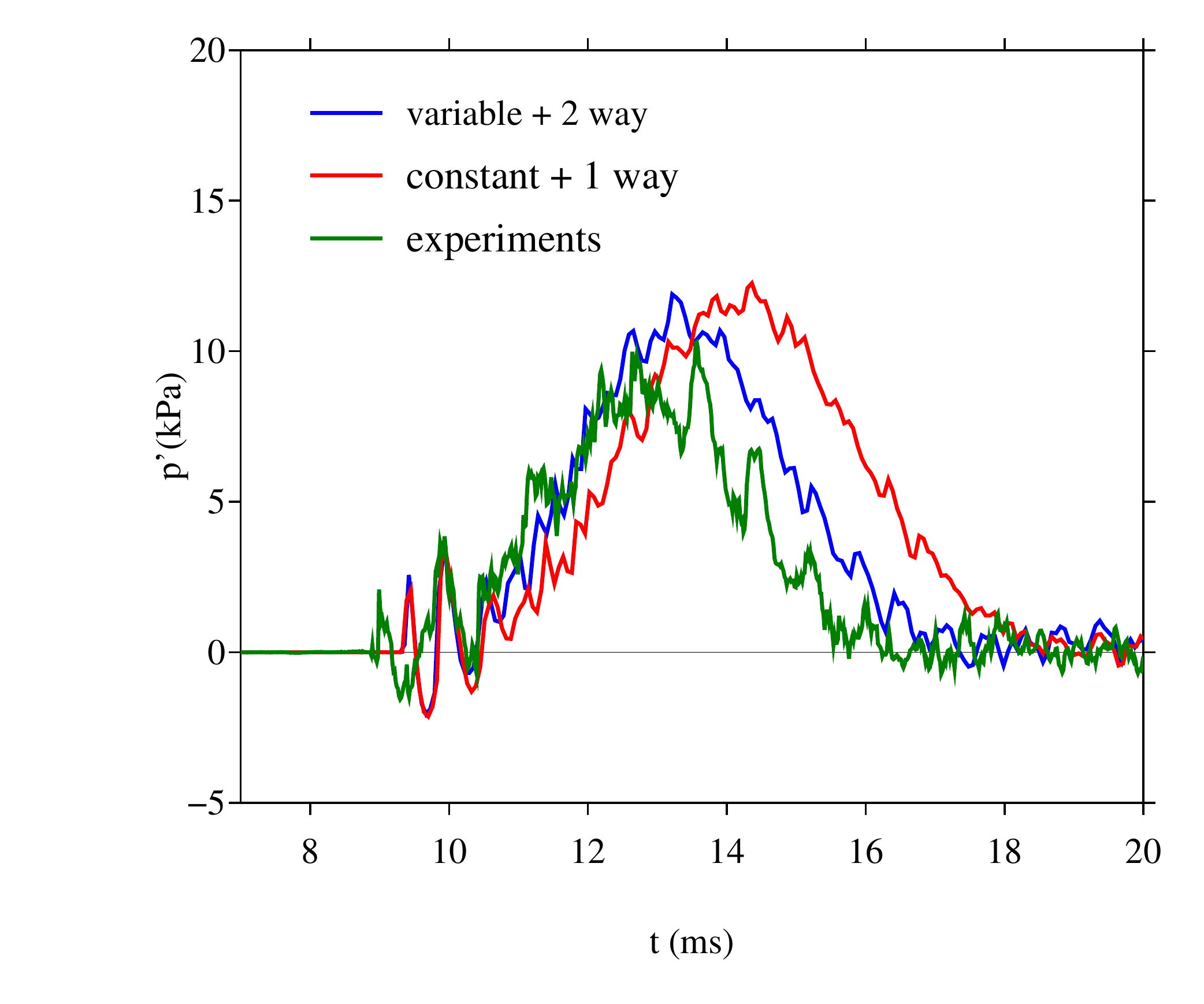} 
\end{tabular}
\end{center}
\vspace{-0.8cm}
\caption{Test1. Time history of the excess pressure $p^{'}$ at $x_r=2.8$ m. Comparison between the models and the experiments. (a): $H=7$ cm; (b): $H=13$ cm.}
\label{FigTest1}
\end{figure}

Fig. \ref{FigTest1} (a-b) compares the excess pressure $p^{'}$, defined in Eq. (\ref{pprime}), simulated by the models or measured experimentally. Two heights are considered: $H=7$ cm (a) and $H=13$ cm (b). The signal is measured at $x_r=2.8$ m. For both models, a good agreement is observed between the simulated data and the experimental results. The best agreement is obtained with the new model, especially for larger resonators heights. It is observed that higher resonators yield smaller waves with a larger support. Moreover, the arrival time of the central peak is around 10 ms in (a) and 14 ms in (b), which indicates that higher resonators generate slower waves, as predicted by the theory \cite{Sugimoto92}.  

  
\subsection{Test 2: amplitude dependence of the velocity}\label{SecTest2}

\begin{figure}[h!]
\begin{center}
\begin{tabular}{cc}
(a) $H=0$ cm & (b) $H=2$ cm\\
\hspace{-1.2cm}
\includegraphics[scale=0.36]{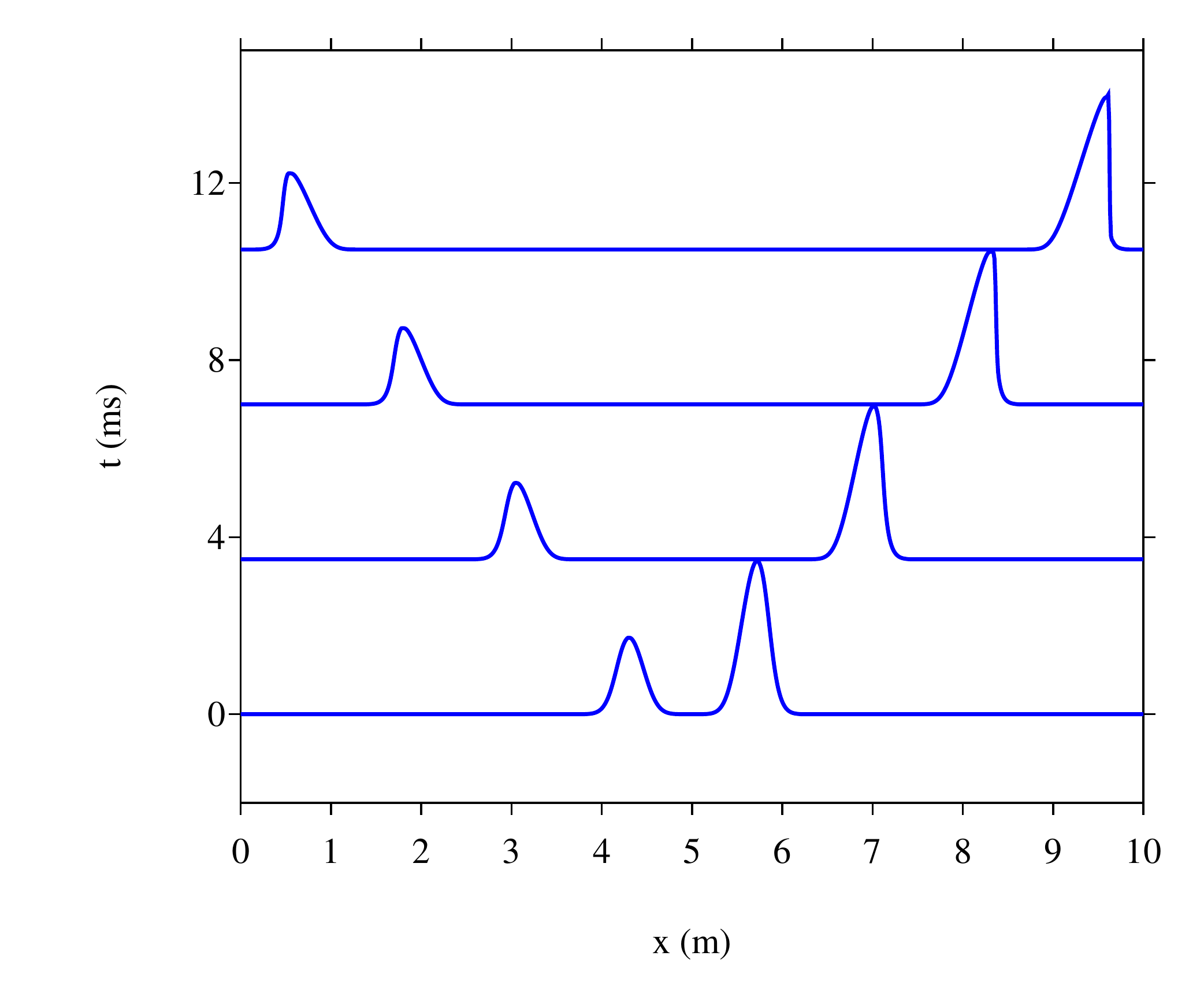} &
\hspace{-1.2cm}
\includegraphics[scale=0.36]{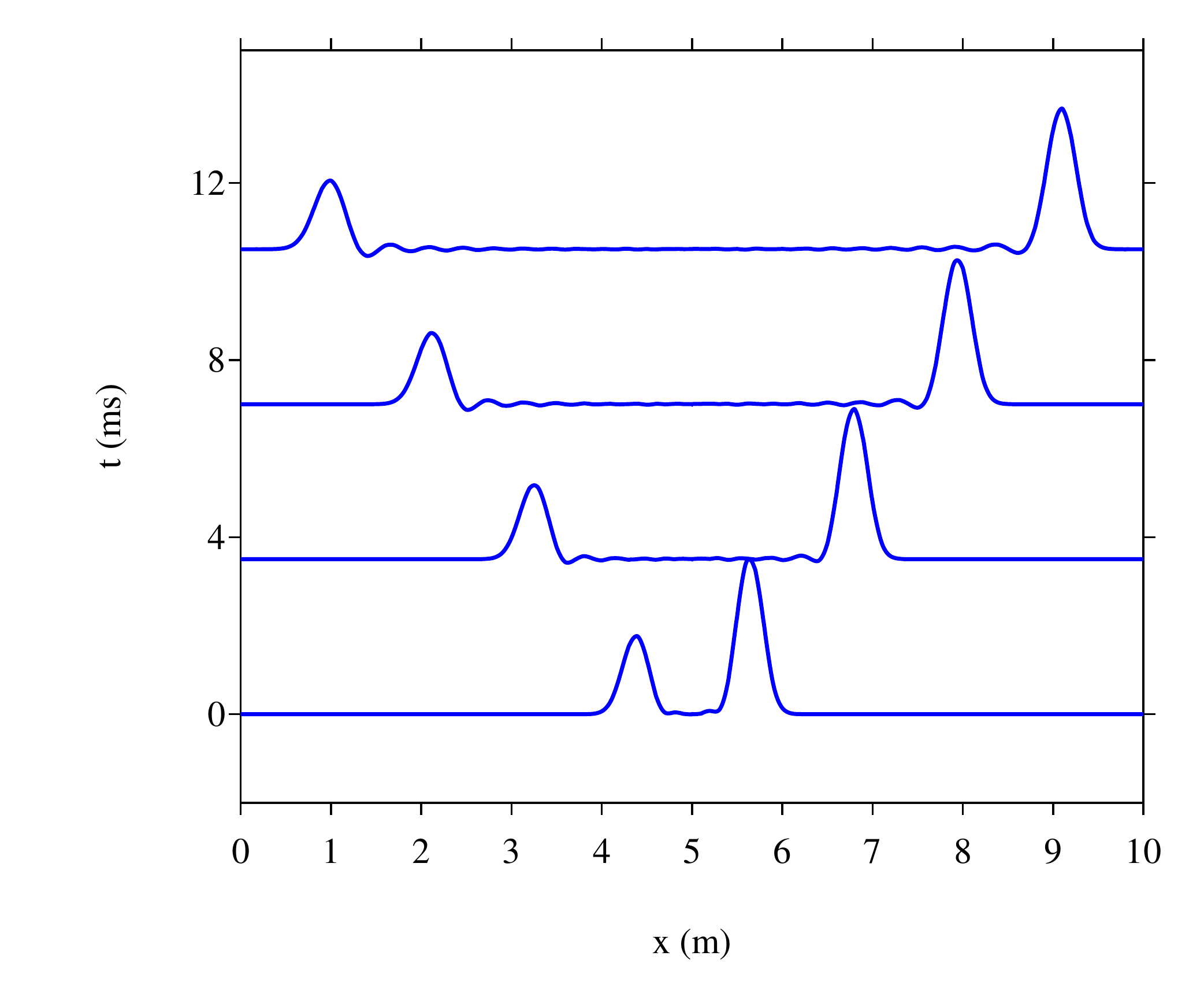} \\
\end{tabular}
\end{center}
\vspace{-0.8cm}
\caption{Test 2. Snapshots of the excess pressure $p^{'}$ generated by a source point at $x=5$ m, at various instants. (a): $H=0$ cm (no resonators); (b): $H=2$ cm.}
\label{FigTest2-sismo}
\end{figure}

From now on, we investigate the properties of the two-way model thanks to numerical experiments. As a second test, we consider the spatial domain $[0,10]$ m, discretized on 5000 grid nodes with $N=4$ memory variables. A time-variable punctual source at $x_s=5$ m is used to generate right-going and left-going waves:
\begin{equation}
u^\pm(x_s,t)=A^\pm\,G(t),
\label{Source}
\end{equation}
where $G$ is a causal Gaussian pulse with $G(0)=0$ and with a central frequency $f_c=650$ Hz. The amplitudes are $A^+=20$ m/s and $A^-=A^+/2=10$ m/s. 

Figure \ref{FigTest2-sismo}(a,b) display snapshots of $p^{'}$, defined in Eq. (\ref{pprime}), at different times, for two resonators heights: $H=0$ cm (a), and $H=2$ cm (b). In the first case without resonators, the viscothermal losses are insufficient to prevent the occurence of shocks in finite time, as analysed theoretically in \cite{Sugimoto91}.
On the contrary, resonators create smooth waves (b).

\begin{figure}[h!]
\begin{center}
\begin{tabular}{c}
\includegraphics[scale=0.40]{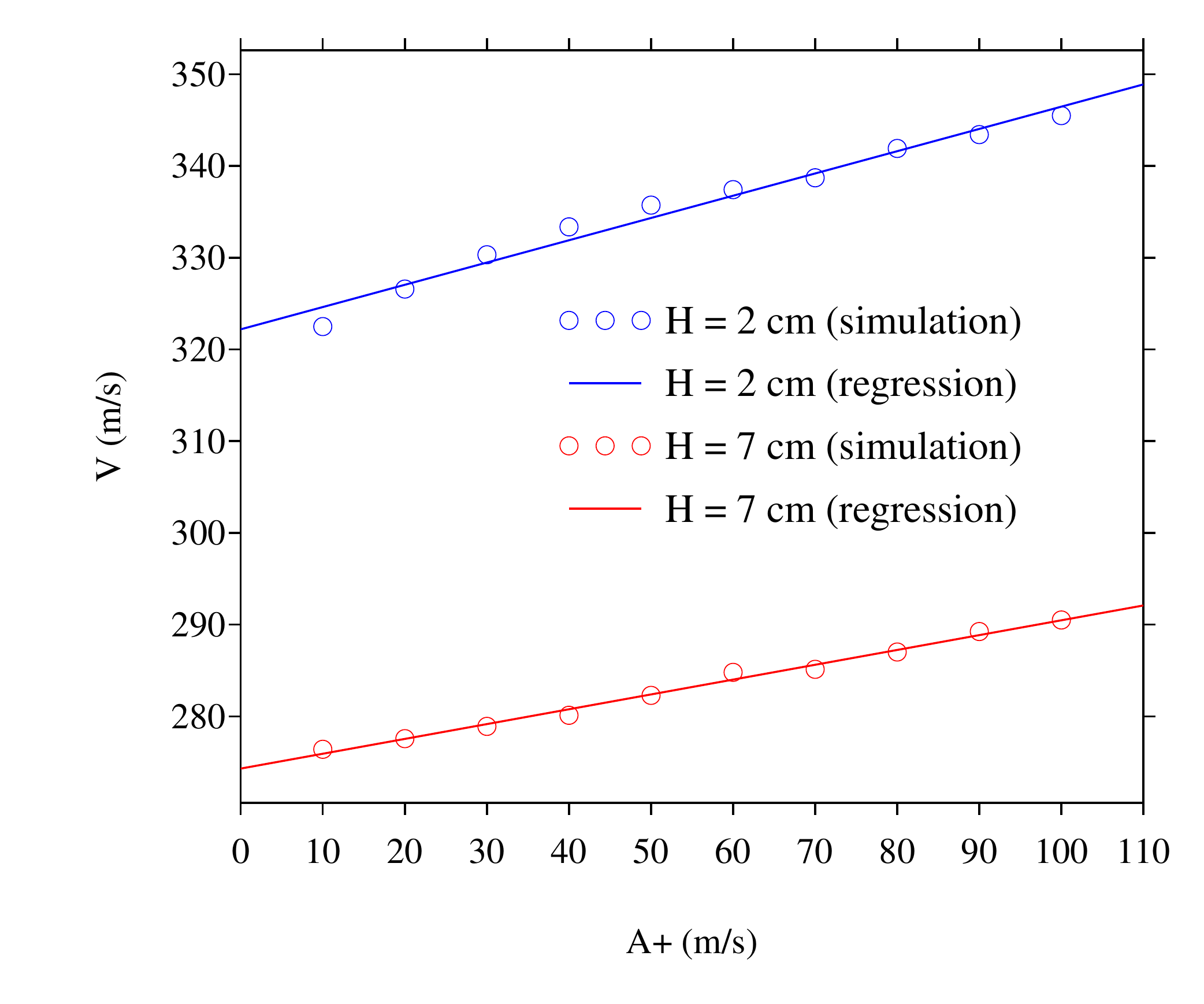}
\end{tabular}
\end{center}
\vspace{-0.8cm}
\caption{Test 2. Velocity of the nonlinear waves versus the forcing amplitude $A^+$ for different heights of the resonators.}
\label{FigTest2-VvsA}
\end{figure}
From Figure \ref{FigTest2-sismo}, one estimates the velocity ${\cal V}$ of the nonlinear waves. Figure \ref{FigTest2-VvsA} shows the results obtained by varying the amplitude $A^+$ of the forcing, from 10 m/s to 100 m/s, and for two heights $H$ of the resonators. It is observed that ${\cal V}$ increases linearly with $A^+$, which is a typical signature of solitary waves. At a given forcing, ${\cal V}$ is also greater for smaller height $H$, as in the first test of section \ref{SecTest1}.

  
\subsection{Test 3: interaction with defects}\label{SecTest3}

\begin{figure}[h!]
\begin{center}
\begin{tabular}{cc}
(a) & (b)\\
\hspace{-1.2cm}
\includegraphics[scale=0.36]{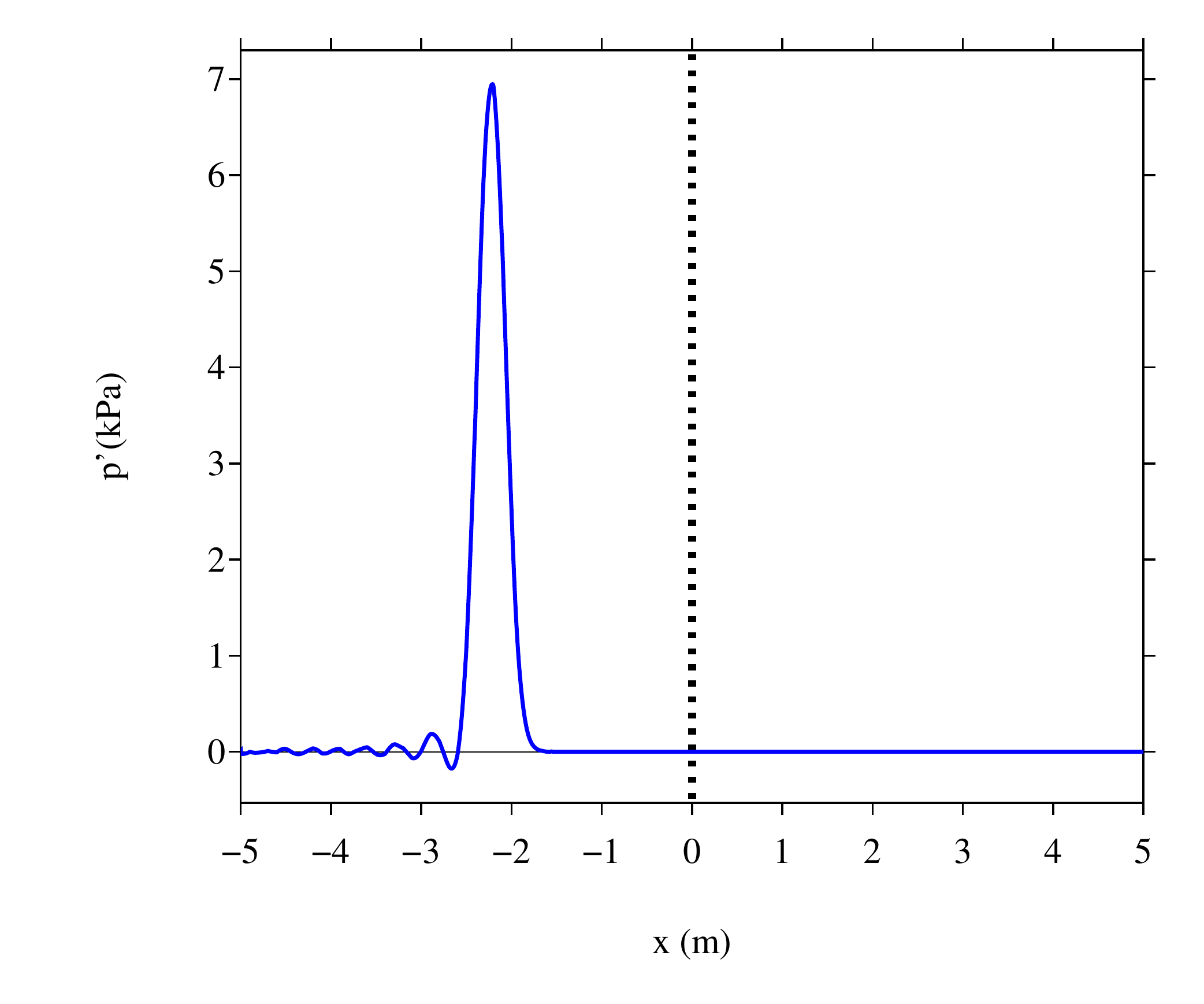} &
\hspace{-1.2cm}
\includegraphics[scale=0.36]{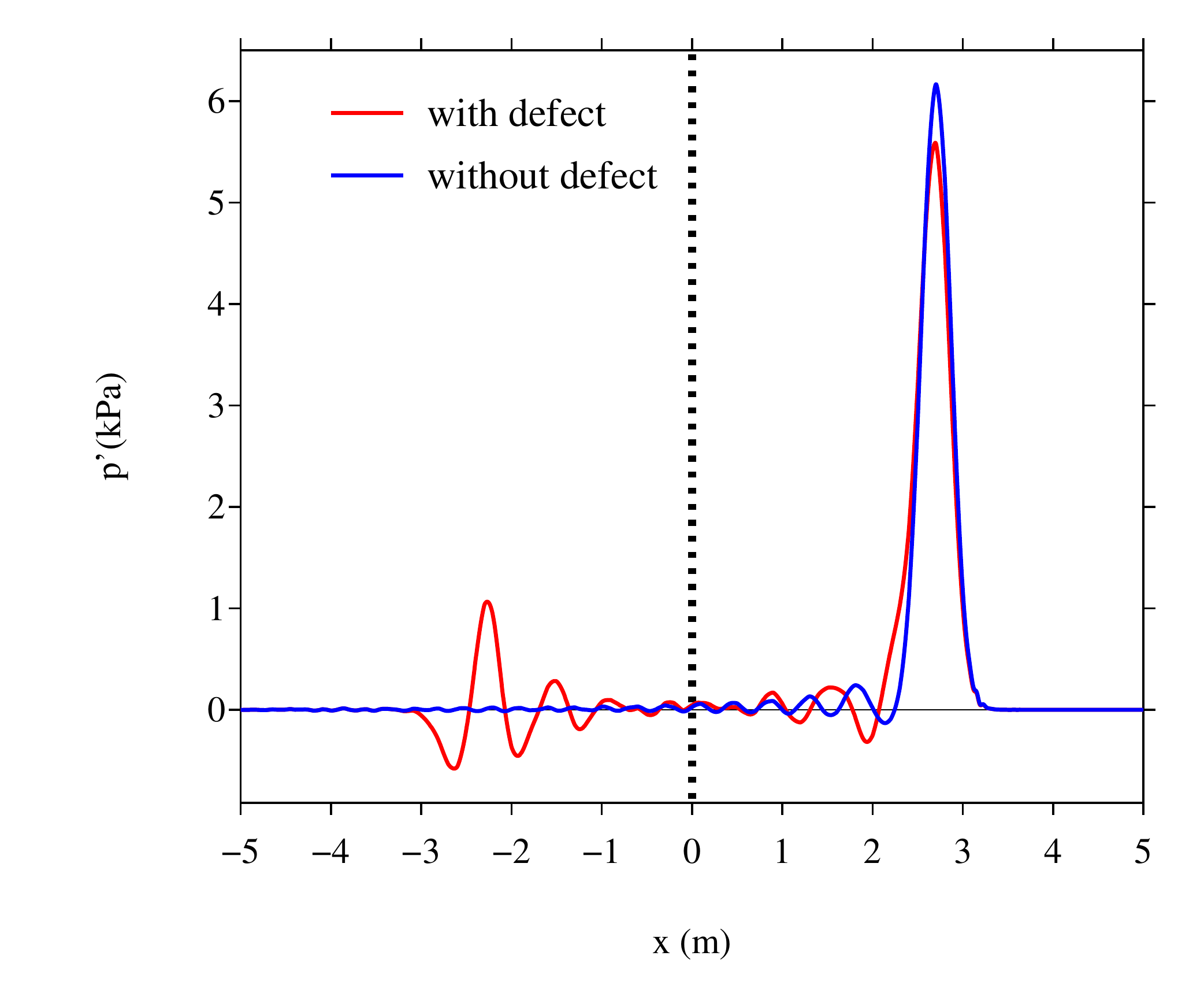} 
\end{tabular}
\end{center}
\vspace{-0.8cm}
\caption{Test 3. Snaphsots of $p^{'}$ at $t=10$ ms (a) and $t=25$ ms (b). All the resonators have the height $H=2$ cm, except at $x=0$ (vertical dotted line), where $H=0.1$ cm.}
\label{FigTest3}
\end{figure}

In this test, the source of right-going waves is localized at the left edge of the computational domain ($x=-5$ m). The amplitude of the forcing is $A^+=20$ m/s; from now on, $A^-=0$ m/s. The height of the resonators is $H=2$ cm, except for the resonator at $x=0$ m, where $H=0.1$ cm. Figure \ref{FigTest3} displays snapshot of $p^{'}$ at different times. The position of the defect is denoted by a vertical dotted line. At $t=10$ ms, the incident wave has not yet crossed the defect (figure \ref{FigTest3}(a)) and exhibits a smooth shape already seen in the second test. 

Figure \ref{FigTest3}(b) shows $p^{'}$ at $t=25$ ms, after the interaction with the defect. The amplitude of the transmitted wave (in red) has been slightly modified, compared to the perfect case without defect (in blue). More important, a left-going wave has been generated by the defect. This effect cannot be predicted by the one-way model with constant coefficients, and it constitutes an original feature of the new model. 

  
\subsection{Test 4: propagation in a random medium}\label{SecTest4}

\begin{figure}[h!]
\begin{center}
\begin{tabular}{cc}
(a) & (b)\\
\hspace{-1.2cm}
\includegraphics[scale=0.36]{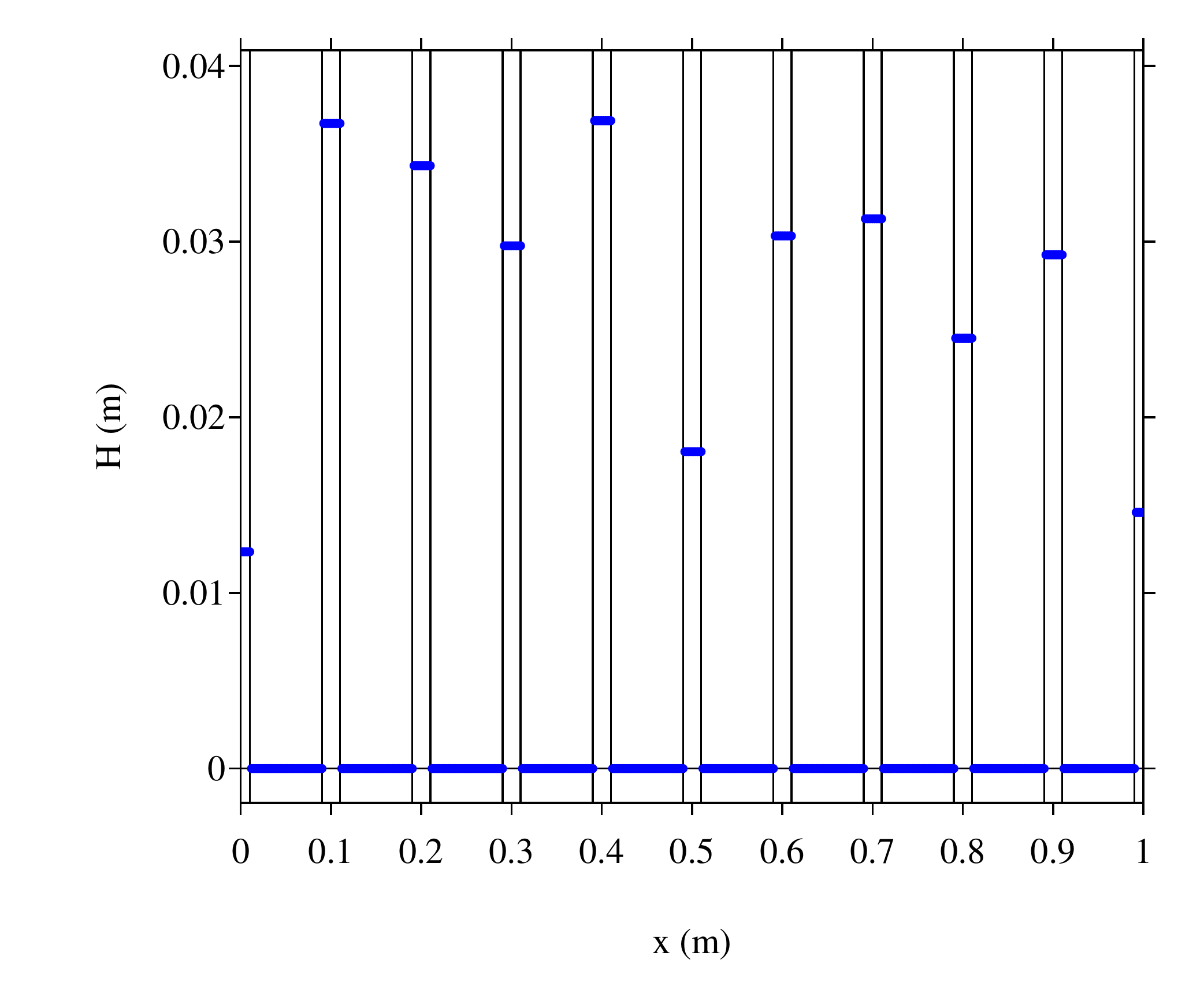} &
\hspace{-1.2cm}
\includegraphics[scale=0.36]{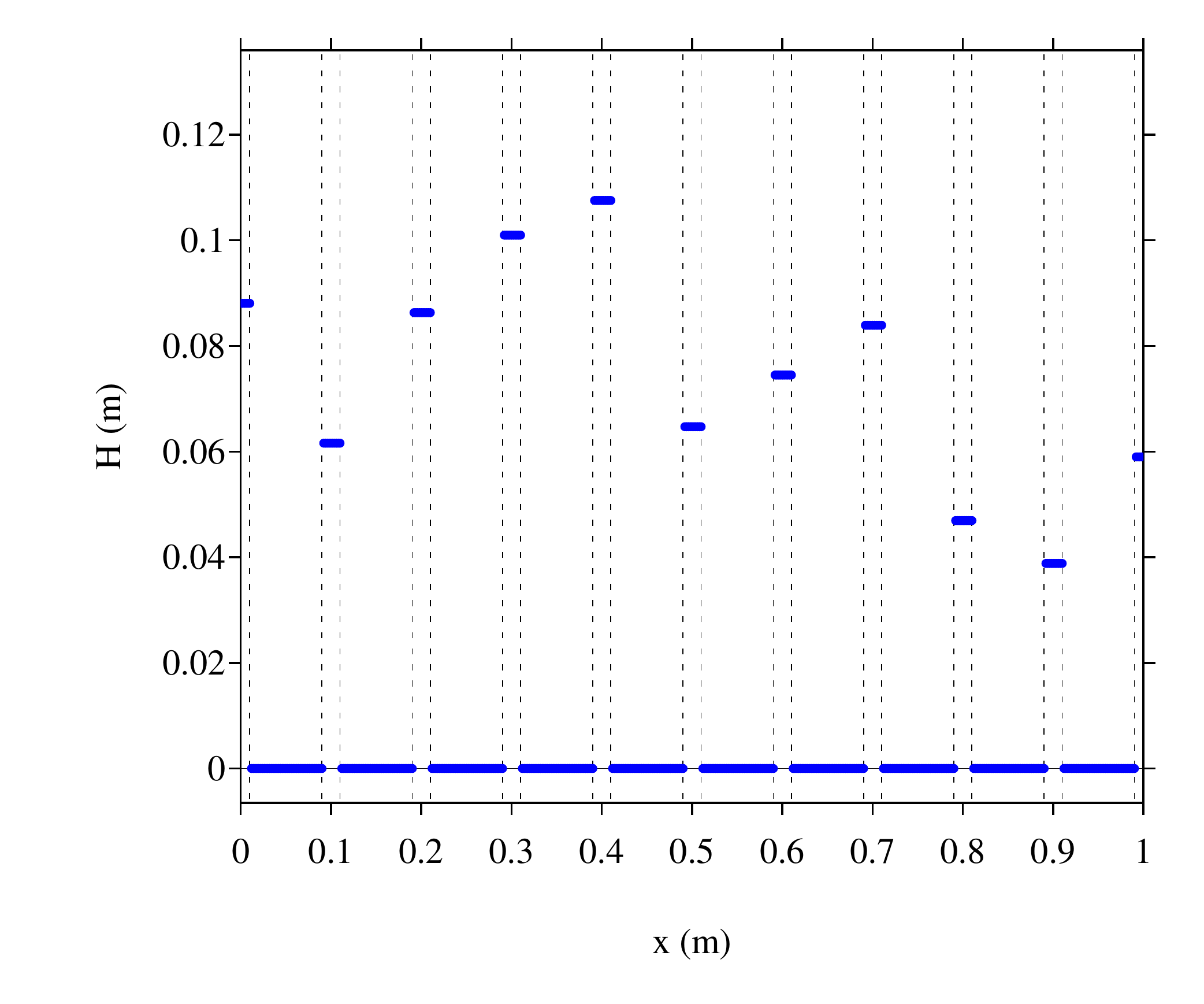} \\
(c) & (d)\\
\hspace{-1.2cm}
\includegraphics[scale=0.36]{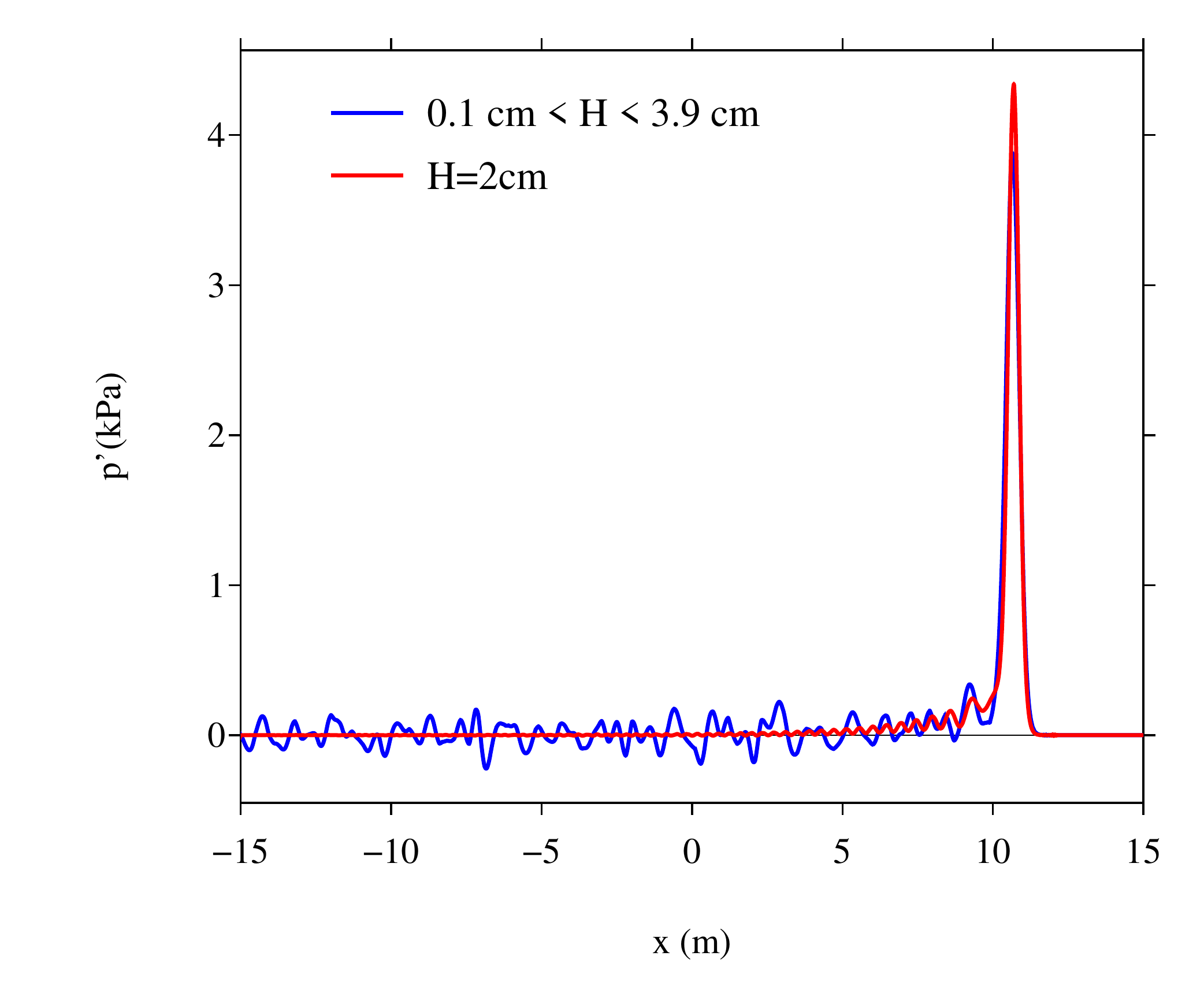} &
\hspace{-1.2cm}
\includegraphics[scale=0.36]{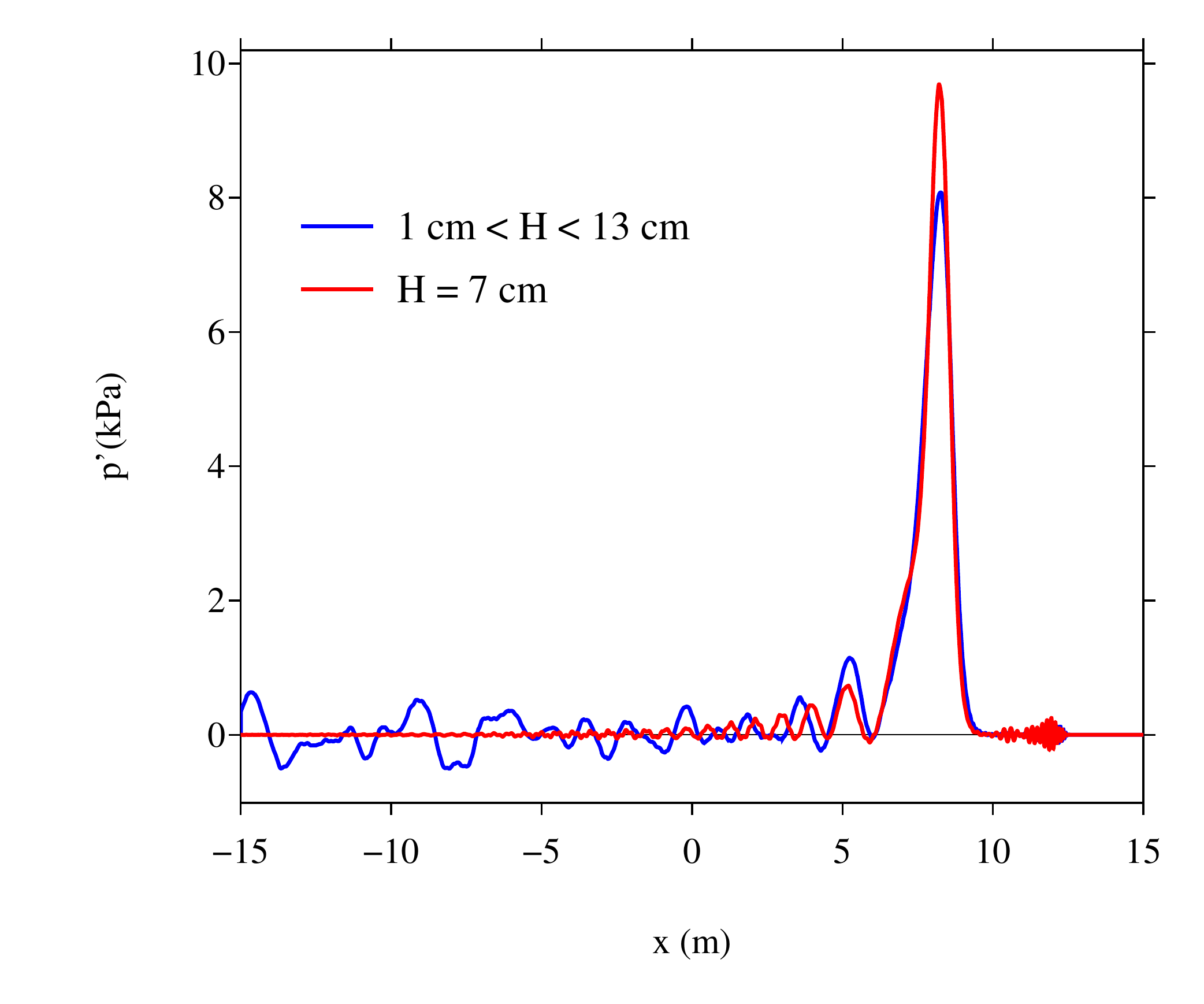} 
\end{tabular}
\end{center}
\vspace{-0.8cm}
\caption{Test 4. Left: $\overline{H}=2$ cm (a-c); right: $\overline{H}=7$ cm (b-d). Top: discretizations of the random heights. The grid nodes are denoted by blue points; the vertical dotted lines denote the positions of the resonators. Bottom: snapshots of $p^{'}$ at $t=80$ ms in the unperturbed medium (red) and in the random medium (blue).}
\label{FigTest4}
\end{figure}

As a fourth and last test, we examine the effect of a random disorder on the propagation of nonlinear waves. The domain $[-15, 15]$ m is discretized on 15000 grid nodes, ensuring 10 grid nodes inside each resonator. The random height of each resonator is uniformly distributed between $H_{\min}$ and $H_{\max}$, which gives rise to random values of the coefficients $e$, $g$, $n$ and $h$ in (\ref{NewModel}). Figure \ref{FigTest4} illustrates two configurations. The first configuration has a mean height $\overline{H}=2$ cm, with $H_{\min}=0.1$ cm and $H_{\max}=3.9$ cm, and a forcing amplitude of right-going waves $A^+=20$ m/s (a,c). The second configuration has a mean height $\overline{H}=7$ cm, with $H_{\min}=1$ cm and $H_{\max}=13$ cm, and a forcing amplitude $A^+=100$ m/s (b,d). A zoom between 0 and 1 m illustrates the values of $H$ in 11 resonators (in blue), denoted by vertical solid lines (a,b).

Figure \ref{FigTest4}(c,d) shows a snapshot of $p^{'}$ at $t=80$ ms without disorder (blue line) and with a random selection of $H$ (red line). With randomness, a coda is observed behind the main wavefront. The amplitude and the location of the peak are only slightly modified, which indicates the robustness of the nonlinear wave when perturbed by some noise. It constitutes an additional signature of solitary waves.


\section{Conclusion}\label{SecConclu}

The goal of this paper was to build a model describing the propagation of nonlinear wave across a variable lattice of Helmholtz resonators. Contrary to previous works \cite{Sugimoto92,Sugimoto04,Richoux15}, the new model takes into account the individual features of each resonator and the backscattering effects. Doing so has enabled us to study the effect of disorder or of an individual defect on the propagation of nonlinear waves. 

The numerical experiments have highlighted various properties of the new model. First, the Helmholtz resonators prevent from the occurence of shocks and yield large-amplitude smooth structures. Moreover, the velocity of waves increases linearly with the amplitude, which confirms the terminology of "acoustic solitary waves". These properties were already included in the original Sugimoto's model \cite{Sugimoto92}, but here a better agreement is obtained with experimental results. Second, diffraction effects have been displayed when a local variation of the height of the resonator is introduced in the lattice. Third and last, robustness to disorder has been observed by performing simulations in random lattices.

This work motivates new experimental investigations. The setup used in \cite{Richoux15} enables to examine whether the effects seen in figures \ref{FigTest3}-\ref{FigTest4} are confirmed experimentally. New theoretical and numerical \cite{xiu} developments to study the propagation in a stochastic medium are also motivated, for instance concerning the robustness of acoustic solitary waves in random media \cite{Garnier07}. Lastly, adequate choices of defects in the lattice could produce localized modes, generalizing the results presented in the linear case in \cite{Sugimoto06}.

\section*{Acknowledgement}

The authors thank O. Richoux who gave access to his experimental measurements. Agn\`es Maurel is also greatly acknowledged for her support. We thank also the anonymous Reviewers for their constructive remarks.


\appendix

\section{Transformation of the fractional integral} \label{SecFractional}

Here we prove (\ref{Dm12}). To simplify, we restrict to a right-way propagation in (\ref{varu}), and we note $u=u^+$ and $p=p^+$:
\begin{equation} \label{varuhomo}
\frac{\partial u}{\partial t}+\frac{\partial}{\partial x}\left(a_0 u + b \ds \frac{u^2}{2}\right) - \ds c \frac{ \partial^{-1/2}}{\partial t^{-1/2}}\frac{\partial u}{\partial x} - d \frac{\partial^2 u}{\partial x^2} = - e (1-2 m p) \ds \frac{\partial p}{\partial t}.
\end{equation}
with the coefficients defined in (\ref{m}) and (\ref{cetedex}). The first step is to write (\ref{varuhomo}) in a dimensionless form. Given a central frequency $f_0$, we define the characteristic wavelength $\lambda=a_0/f_0$. We introduce the non-dimensional quantities, indexed by a tilde: $x=\lambda \tilde{x}$, $t=\tilde{t}/f_0$, $u = u_0 \tilde{u}$ and $p = p_0 \tilde{p}$, where $u_0$ is the characteristic velocity, that will be determined later.
We get:
\begin{equation} \label{sansdim}
\frac{\partial \tilde{u}}{\partial \tilde{t}} +\frac{\partial}{\partial \tilde{x}}\left(\tilde{u} + \tilde{b} \ds \frac{\tilde{u}^2}{2}\right) - \ds \tilde{c} \frac{\partial^{-1/2}}{\partial \tilde{t}^{-1/2}}\frac{\partial \tilde{u}}{\partial \tilde{x}} - \tilde{d} \frac{\partial^2 \tilde{u}}{\partial \tilde{x}^2} = - \tilde{e} (1-2 \tilde{m} \tilde{p}) \ds \frac{\partial \tilde{p}}{\partial \tilde{t}},
\end{equation}
with
\begin{equation} \label{}
\tilde{b} = M b = M \frac{\gamma+1}{2 },\, \tilde{c}=\frac{c}{a_0} \sqrt{\frac{\lambda}{a_0}}
,\, \tilde{d}=\frac{d}{a_0 \lambda},\, \tilde{e}=\frac{e p_0}{u_0}=\frac{V}{2 A D \gamma M},\, \tilde{m}=m p_0=\frac{\gamma-1}{2 \gamma}.
\end{equation}
$M=u_0/c_0$ is the characteristic Mach number. The typical value of $M$ is obtained by equalizing $\tilde{b}$ and $\tilde{e}$:
\begin{equation} \label{Mvalue}
\tilde{b} = \tilde{e} \Leftrightarrow M  = \sqrt{ \frac{V}{A D \gamma (\gamma+1)} }.
\end{equation} 
To find an approximation of (\ref{sansdim}), we introduce the small parameter $\varepsilon = \max(\tilde{b},\tilde{c},\tilde{d},\tilde{e})$. Then we define new quantities, defined with hats, by the relation $v=\varepsilon \hat{v},$ with $v=b$, $c$, $d$ or $e$, such that all the quantities with a hat are at most equal to one. Then, starting from (\ref{sansdim}), we get:
\begin{eqnarray*}
\frac{\partial \tilde{u}}{\partial \tilde{x}} &=& - \frac{\partial \tilde{u}}{\partial \tilde{t}} - \varepsilon \hat{b} \tilde{u} \ds \frac{\partial \tilde{u}}{\partial \tilde{x}} + \varepsilon \hat{d} \frac{\partial^2 \tilde{u}}{ \partial \tilde{x}^2} - \varepsilon \hat{e} (1-2 \tilde{m} \tilde{p}) \ds \frac{\textstyle \partial \tilde{p}}{\partial \tilde{t}} + \varepsilon \hat{c} \frac{\partial^{-1/2}}{\partial \tilde{t}^{-1/2}}\frac{\partial \tilde{u}}{ \partial \tilde{x}},\\
[8pt]
&=& -\frac{\partial \tilde{u}}{\partial \tilde{t}} - \varepsilon \hat{b} \tilde{u} \ds \frac{\partial \tilde{u}}{\partial \tilde{x}} + \varepsilon \hat{d} \frac{\partial^2 \tilde{u}}{\partial \tilde{x}^2} - \varepsilon \hat{e} (1-2 \tilde{m} \tilde{p}) \ds \frac{\partial \tilde{p}}{\partial \tilde{t}}\\
[8pt]
&+& \varepsilon \hat{c} \frac{\partial^{-1/2}}{\partial \tilde{t}^{-1/2}} \left( - \frac{\partial \tilde{u}}{\partial \tilde{t}} - \varepsilon \hat{b} \tilde{u} \ds \frac{\partial \tilde{u}}{\partial \tilde{x}} + \varepsilon \hat{d} \frac{\partial^2 \tilde{u}}{\partial \tilde{x}^2} - \varepsilon \hat{e} (1-2 \tilde{m} \tilde{p}) \ds \frac{\partial \tilde{p}}{\partial \tilde{t}} + \varepsilon \hat{c} \frac{\partial^{-1/2}}{\partial \tilde{t}^{-1/2}}\frac{\partial \tilde{u}}{\partial \tilde{x}} \right), \\
[8pt]
&=& - \frac{\partial \tilde{u}}{\partial \tilde{t}} - \varepsilon \hat{b} \tilde{u} \ds \frac{\partial \tilde{u}}{\partial \tilde{x}} + \varepsilon \hat{d} \frac{\partial^2 \tilde{u}}{\partial \tilde{x}^2} - \varepsilon \hat{e} (1-2 \tilde{m} \tilde{p}) \ds \frac{\partial \tilde{p}}{\partial \tilde{t}} - \varepsilon \hat{c} \frac{\partial^{1/2} \tilde{u}}{\partial \tilde{t}^{1/2}} + O(\varepsilon^2),
\end{eqnarray*}
where we have used
$$
\frac{\partial^{1/2} \tilde{u}}{\partial \tilde{t}^{1/2}} = \frac{\partial^{-1/2}}{\partial \tilde{t}^{-1/2}}\frac{\partial \tilde{u}}{\partial \tilde{t}}.
$$
Therefore (\ref{sansdim}) can be approximated by:
\begin{equation} \label{sansdimder}
\frac{\partial \tilde{u}}{\partial \tilde{t}} + \frac{\partial}{\partial \tilde{x}}\left(\tilde{u} + \tilde{b} \ds \frac{\tilde{u}^2}{2}\right) - \ds \tilde{c} \frac{\partial^{1/2} \tilde{u}}{\partial \tilde{t}^{1/2}} - \tilde{d} \frac{\partial^2 \tilde{u}}{\partial \tilde{x}^2} = - \tilde{e} (1-2 \tilde{m} \tilde{p}) \ds \frac{\textstyle \partial \tilde{p}}{\partial \tilde{t}},
\end{equation}
where neglected terms are small, of order $\varepsilon^2$. Using the values of Table \ref{TabParam}, $M=0.21$ is obtained from (\ref{Mvalue}), in agreement with the experiments made in \cite{Richoux15}, for $f_0=500$ Hz. From this Mach value, we deduce finally
\begin{equation} \label{}
\tilde{b} = 0.22 = \tilde{e},\quad \tilde{c} = 0.010,\quad \tilde{d} =8.24 10^{-8},\quad \tilde{m} = 0.14.
\end{equation}
Coming back to real quantities yields the first two equations in (\ref{NewModel}). 


\section{Comparison with Sugimoto's model}\label{SecSugi}

Here we prove that the original model of Sugimoto \cite{Sugimoto92} can be recovered from the new model (\ref{NewModel}). In this aim, we introduce three ingredients:
\begin{itemize}
\item a restriction to only one-way propagations: for right-going waves ($u^-=0$), we note $p^+$ the associated pressure $p$. Similarly, for left-going waves ($u^+=0$), we note $p^-$ the pressure $p$;
\item in the tube, the coefficient of adiabatic nonlinearity $2mp$ is neglected;
\item an averaged description of the geometry is used: for identical resonators of volume $V$, a continuous approximation of the tube geometry is introduced and a mean value of the flux is used:
\begin{equation} \label{}
\bar{F}^\pm = \frac{1}{D} \int_{x=0}^D F^\pm(x) dx.
\end{equation}
\end{itemize}
Using (\ref{funccete}) and (\ref{coeffbis}), one deduces the mean value
\begin{equation} \label{emoy}
\bar{e} = \frac{1}{D} \int_{x=0}^D e(x) dx = e_0 \frac{B}{2D} = \frac{\textstyle V}{\textstyle 2\,\rho_0\,a_0\,A\,D}.
\end{equation}
Also, thanks to the approximation for $R \gg r$:
\begin{equation} \label{} \arcsin \left( \frac{\sqrt{r^2-x^2}}{R} \right) \simeq \frac{\sqrt{r^2-x^2}}{R}.\end{equation}
Using (\ref{funccete}) and (\ref{coeffbis}), we obtain the mean value
\begin{equation} \label{cmoy}
\bar{c} = \frac{1}{D} \int_{x=0}^D c(x) dx = c_0 \left( 1 - \frac{r^2}{2 R D} \right) = C \, a_0 \sqrt{\nu} \left( \frac{1}{R} - \frac{B}{2 A D} \right) = C \, a_0 \sqrt{\nu} \frac{1}{R^*},
\end{equation}
with
\begin{equation} \label{Rstar}
\frac{1}{R^*} = \frac{1}{R} - \frac{B}{2 A D}.
\end{equation}
Using these two ingredients, the system (\ref{NewModel}) degenerates in two families of uncoupled equations: 
\begin{subnumcases}{\label{sugi}}
\ds \frac{\partial u}{\partial t}^\pm + \frac{\partial}{\partial x} \left(\pm a_0 u^\pm + b \ds \frac{\textstyle (u^\pm)^2}{2} \right) \ds \mp \bar{c} \frac{\partial^{-1/2}}{\partial t^{-1/2}} \frac{\partial u}{\partial x} ^\pm - d \frac{\partial^2 u}{\partial x^2}^\pm = \mp \bar{e} \ds \frac{\partial p}{\partial t}^\pm, \label{sugiu} \\
[8pt]
\ds \frac{\partial^2 p}{\partial t^2}^\pm + f \frac{\partial^{3/2} p}{\partial t^{3/2}}^\pm + g p^\pm - m \frac{\partial^2 (p^\pm)^2}{\partial t^2}+n\left|\frac{\partial p}{\partial t}^\pm \right|\,\frac{\partial p}{\partial t}^\pm = \pm hu^\pm, \label{sugip}
\end{subnumcases}
which corresponds to the model proposed in \cite{Sugimoto92,Sugimoto04} and studied in \cite{Richoux15}.


\section{Energy balance} \label{SecNRJproof}

Here we prove the result \ref{energie}. The first step is to rewrite the system (\ref{EDPbis}), eliminating the term $2 m |p| \ll 1$ in (\ref{EDP1bis}) and using (\ref{ODED12}), to get the following system:
\begin{subnumcases}{\label{EDP}}
\ds
\frac{\partial u^\pm}{\partial t}+\frac{\partial}{\partial x}\left(\pm au^\pm + b\frac{(u^\pm)^2}{2}\right)= - \frac{c}{a_0} \sum_{\ell=1}^N\mu_{\ell}\frac{\partial \varphi_\ell^\pm}{\partial t}+d\frac{\partial^2 u^\pm}{\partial x^2}\mp eq,\label{EDP1}\\
\ds
\frac{ \partial p}{\textstyle \partial t}=q,\label{EDP2}\\
\ds
\frac{\partial q}{\partial t}=h(u^+-u^-)-gp-f\sum_{\ell=1}^N\mu_{\ell}\frac{\partial \xi_{\ell}}{\partial t}+m\frac{ \partial^2 p^2}{\partial t^2}-n\,\left|q\right|\,q,\label{EDP3}\\
\ds
\frac{\partial \varphi^\pm_{\ell}}{\partial t}=-\theta_{\ell}^2\varphi^\pm_{\ell}+\frac{2}{\pi}u^\pm,\hspace{1.5cm} \ell=1,\cdots, N,\label{EDP5}\\
[6pt]
\ds
\frac{\partial \xi_{\ell}}{\partial t}=-\theta_{\ell}^2\xi_{\ell}+\frac{2}{\pi}q,\hspace{2.1cm} \ell=1,\cdots, N.\label{EDP6}
\end{subnumcases}
From (\ref{EDP5}) and (\ref{EDP6}), it follows
\begin{equation}
u^\pm=\frac{\pi}{2}\left(\frac{\partial \varphi^\pm_{\ell}}{\partial t}+\theta_{\ell}^2\,\varphi^\pm_{\ell}\right),\hspace{1cm}q=\frac{\pi}{2}\left(\frac{\partial \xi_{\ell}}{\partial t}+\theta_{\ell}^2\,\xi_{\ell}\right).
\label{upm}
\end{equation}
In (\ref{EDP3}), the derivative of $p^2$ is modified, using
\begin{equation}
\frac{\partial^2}{\partial t^2}(p^2)\,q=\frac{\partial}{\partial t}(p\,q^2)+q^3.
\label{d2p2}
\end{equation}
This relation is easy to check by expanding both sides of the equality. Next we multiply (\ref{EDP3}) by $q$, and using (\ref{EDP2}), (\ref{upm}) and (\ref{d2p2}) leads to
\begin{equation}
\begin{array}{lll}
\ds q\frac{\partial q}{\partial t} &=& \ds h(u^+-u^-)\,q-g\,p\,q-f\,\sum_{\ell=1}^N\mu_{\ell}\frac{\partial \xi_{\ell}}{\partial t}\,q+m\frac{ \partial^2 p^2}{\partial t^2}\,q-n\,\left|q\right|\,q^2,\\
[8pt]
&=& \ds h(u^+-u^-)\,q - g\,p\frac{\partial p}{\partial t}-\frac{\pi}{2}f\sum_{\ell=1}^N\mu_{\ell}\left(\frac{\partial \xi_{\ell}}{\partial t}+\theta_{\ell}^2\,\xi_{\ell}\right)\frac{\partial \xi_{\ell}}{\partial t}+m\frac{\partial}{\partial t}(p\,q^2)\\
[6pt]
&& \ds -nq^2\left(|q|-\frac{m}{n}q\right).
\end{array}
\end{equation}
Therefore, we get
\begin{equation}
\begin{array}{lll}
\ds (u^+-u^-) q &=& \displaystyle \frac{1}{2} \frac{\partial}{\partial t}\left(\frac{g}{h}p^2+\frac{1}{h}(1-2\,m\,p)\,q^2+\frac{\pi}{2}\frac{f}{h}\sum_{\ell=1}^N\mu_\ell\,\theta_\ell^2\,\xi_\ell^2\right)\\
[6pt]
&& \ds + \frac{\pi}{2}\,\frac{f}{h}\sum_{\ell=1}^N\mu_\ell\left(\frac{\partial \xi_{\ell}}{\partial t}\right)^2+\frac{n}{h}\,q^2\left(|q|-\frac{m}{n}q\right).
\end{array}
\label{Up-Um}
\end{equation}
Besides, (\ref{EDP1}) is multiplied by $u^\pm$ and integrated in space. After summation and integration by parts (the data are compactly supported), we get
\begin{equation}
\begin{array}{l}
\ds \int_\mathbb{R}\left(u^+\,\frac{\partial u^+}{\partial t}+u^-\,\frac{\partial u^-}{\partial t}\right)\,dx = \displaystyle-\int_\mathbb{R} \frac{c}{a_0} \sum_{\ell=1}^N \mu_\ell \left( u^+ \frac{\partial \varphi^+_{\ell}}{\partial t}+u^-\frac{\partial \varphi^-_{\ell}}{\partial t}\right)\,dx\\
\\
\hspace{1cm} \ds -\int_\mathbb{R}d\left(\left(\frac{\partial u^+}{\partial x}\right)^2 + \left(\frac{\partial u^-}{\partial x}\right)^2\right)dx-\int_\mathbb{R}e\,(u^+-u^-)\,q\,dx.
\end{array}
\end{equation}
Thanks to the relations (\ref{upm}) and (\ref{Up-Um}), the previous equation is simplified and the conclusion follows.


\end{document}